\documentclass[12pt]{article}
\pdfoutput=1

\usepackage{jheppub}
\usepackage{amsmath}
\usepackage{amssymb}
\usepackage{graphicx}
\usepackage{xspace}
\usepackage{subfig}

\usepackage{enumitem}

\usepackage{color}
\definecolor{comment}{rgb}{0,0.3,0}
\definecolor{identifier}{rgb}{0.0,0,0.3}
\usepackage{listings}
\newcommand{\ttt}[1]{{\small\texttt{#1}}}

\makeatletter
\g@addto@macro\bfseries{\boldmath}
\makeatother

\usepackage{array}
\newcolumntype{L}[1]{>{\raggedright\let\newline\\\arraybackslash\hspace{0pt}}m{#1}}
\newcolumntype{C}[1]{>{\centering\let\newline\\\arraybackslash\hspace{0pt}}m{#1}}
\newcolumntype{R}[1]{>{\raggedleft\let\newline\\\arraybackslash\hspace{0pt}}m{#1}}

\usepackage{color}
\definecolor{darkgreen}{rgb}{0,0.5,0}
\definecolor{darkblue}{rgb}{0,0,0.7}
\definecolor{darkred}{rgb}{0.75,0,0.0}
\definecolor{darkorange}{rgb}{0.8,0.4,0.0}

\newcommand{\arXiv}[1]{\href{http://arxiv.org/abs/#1}{arXiv:#1}}
\newcommand{\hepph}[1]{\href{http://arxiv.org/abs/hep-ph/#1}{hep-ph/#1}}

\newcommand{\ie}{i.e.\ }
\newcommand{\eg}{e.g.\ }
\newcommand{\order}[1]{{\cal O}\left(#1\right)}
\newcommand{\as}{\alpha_s}

\newcommand{\GeV}{\,\text{GeV}}
\newcommand{\TeV}{\,\text{TeV}}

\newcommand{\zcut}{\ensuremath{z_{\text{cut}}}\xspace}
\newcommand{\zetacut}{\ensuremath{\zeta_{\text{cut}}}\xspace}

\newcommand{\fulljet}{full\xspace}
\newcommand{\taucut}{\tau_{\text{cut}}}

\newcommand{\tauP}{\tau_{21}^\text{\fulljet}}
\newcommand{\tauT}{\tau_{21}^\text{tagged}}
\newcommand{\tauD}{\tau_{21}^\text{dichroic}}

\newcommand{\tauPG}{\tau_{21,\text{groomed}}^\text{\fulljet}}
\newcommand{\tauTG}{\tau_{21,\text{groomed}}^\text{tagged}}
\newcommand{\tauDG}{\tau_{21,\text{groomed}}^\text{dichroic}}

\preprint{CERN-TH/2016-251}

\title{Dichroic subjettiness ratios to distinguish colour flows in boosted boson tagging}
\author[a,*]{Gavin P. Salam,\note[*]{On leave from CNRS, UMR 7589, LPTHE, F-75005, Paris, France}}
\author[b]{Lais Schunk}
\author[b]{and Gregory Soyez}
\affiliation[a]{CERN, Theoretical Physics Department, CH-1211
  Geneva 23, Switzerland}
\affiliation[b]{IPhT, CEA Saclay, CNRS UMR 3681, F-91191 Gif-sur-Yvette cedex, France}

\emailAdd{gavin.salam@cern.ch}
\emailAdd{lais.sarem-schunk@cea.fr}
\emailAdd{gregory.soyez@cea.fr}


\abstract{
  $N$-subjettiness ratios are in wide use for tagging heavy boosted
  objects,
  in particular the ratio of 2-subjettiness to 1-subjettiness for
  tagging boosted electroweak bosons.
  In this article we introduce a new, \emph{dichroic} ratio, which uses
  different regions of a jet to determine the two subjettiness
  measures, emphasising the hard substructure for the 1-subjettiness
  and the full colour radiation pattern for the 2-subjettiness.
  Relative to existing $N$-subjettiness ratios, the dichroic
  extension, combined with SoftDrop (pre-)grooming, makes it possible to
  increase the ultimate signal significance by about $25\%$ (for
  $2\,\text{TeV}$ jets), or to reduce non-perturbative effects by a
  factor of $2{-}3$ at $50\%$ signal efficiency while maintaining
  comparable background rejection.
  We motivate the dichroic approach through the study of Lund
  diagrams, supplemented with resummed analytical calculations.  
}

\begin{document}

\maketitle

\section{Introduction}\label{sec:intro} 

With the increasingly high-energy scales probed by the Large Hadron
Collider (LHC), massive electroweak bosons ($H/Z/W$) and top quarks are often
produced with a transverse momentum much larger than their mass.
In this {\it boosted} regime, when they decay hadronically, they are
reconstructed as single jets that have to be separated from the much
more common quark- and gluon-initiated jets. 
Over the past few years, several techniques relying on {\it jet
  substructure}, \ie on the internal dynamical properties of jets,
have been devised in order to achieve this task. These techniques are
now routinely used in LHC analyses and new-physics searches.

There are three common families of methods used to separate boosted
heavy objects from standard QCD jets:
(i) {\it taggers}, which impose that a jet contain two hard cores (or
three for a top-quark), a situation more common in signal jets than in
QCD jets which are dominated by soft-gluon radiation; an increasingly
widespread technique for tagging is the modified MassDrop tagger
(mMDT)~\cite{Butterworth:2008iy,Dasgupta:2013ihk} and its
generalisation, SoftDrop~\cite{Larkoski:2014wba}, which will be our
chosen tools here;
(ii) {\it radiation constraints}, which constrain soft-gluon radiation
inside jets, expected to be larger in QCD jets than in colourless
weak-boson decays; a widespread way of applying radiation constraints
is to cut on jet shapes, for example the ratio of $\tau_2/\tau_1$,
where $\tau_N$ is the
$N$-subjettiness~\cite{Thaler:2010tr,Kim:2010uj,Thaler:2011gf}.
(iii) {\it groomers}, which clean the fat jets of soft-and-large-angle
radiation, often dominated by the Underlying Event and pileup, hence
ensuring a better mass resolution.

To reach a large discriminating power, it is helpful to combine
several of these techniques. Since taggers and groomers share many
similarities, one often starts by applying a tagger/groomer and then
imposes a cut on the value of a jet shape computed on that tagged/groomed
jet. 
Finally, one selects jets with a (groomed) mass close-enough to the
weak boson mass.

In this paper, we introduce the concept of ``dichroic'' subjettiness
ratios for applying radiation constraints.
Starting from an object in which two hard prongs have been identified
(``tagged''), the dichroic variant of subjettiness differs from
standard subjettiness ratios because it uses different (sub)jets for
the numerator and denominator of the $\tau_2/\tau_1$ ratio.
The reason for calling this ``dichroic'' is that the two different
(sub)jets that are used are dominated by two distinct colour flows.
In particular we will use a large jet for calculating $\tau_2$ and a
smaller, tagged subjet for $\tau_1$.
Calculating $\tau_2$ on the large jet provides substantial
sensitivity to the different colour flows of signal (colour singlet
when viewed at large angles) and background (colour triplet for a
quark-jet or octet for a gluon-jet).
Calculating $\tau_1$ on the tagged subjet ensures that it is not
substantially affected by the overall colour flow of the large jet,
but rather is governed essentially by the invariant mass of the
two-prong structure found by the tagger.
The resulting dichroic $\tau_2/\tau_1$ ratio gives enhanced
performance compared to existing uses of $N$-subjettiness, which adopt
the same (sub)jet for numerator and denominator~(see \eg
\cite{Behr:2015oqq,Aad:2015typ,Khachatryan:2016cfa,Dolen:2016kst,Khachatryan:2016mdm}
for recent examples).

Performance of radiation-based discrimination involves two criteria:
the ability to distinguish signals from backgrounds and the robustness
of that discrimination, notably its insensitivity to non-perturbative
effects. 
As discussed already in~\cite{Dasgupta:2015lxh}, these two criteria
are often in tension, because the region of large-angle soft
kinematics on one hand provides substantial discrimination power, but
is also the region where the Underlying Event and hadronisation have
the largest impact.
A point central in our discussion will be the trade-off between these
aspects.
To reduce the tension between discrimination power and perturbative
robustness we will show how the dichroic subjettiness ratio can be
used in combination not just with tagging but also a separate
(pre-)grooming step.

\section{Setup and useful tools for discussion}\label{sec:setup}

Before introducing the dichroic tools in Section~\ref{sec:new}, let us 
first discuss the individual building blocks used in our new
combination and introduce a simple framework to facilitate the
discussion of the underlying physics and expected performance.

\subsection{A tagger, a groomer and a jet shape}\label{sec:methods}

We will concentrate on the modified
MassDrop tagger, used here
as a tagger, $N$-subjettiness as a radiation-constraining jet
shape, and
SoftDrop as a groomer.
These are all common choices in the literature, but we believe that our
generic strategy can be extended to other combinations if needed. 
To ease the physics discussion below, let us briefly recall how these
methods are defined.

The modified MassDrop tagger and SoftDrop both start by
reclustering the jet with the Cambridge/Aachen algorithm. They then
recursively undo the last step of the clustering, splitting the
current jet $j$ into two subjets $j_1$ and $j_2$. The procedures then
stop if the splitting is symmetric enough, \ie if
\begin{equation}\label{eq:softdrop-condition}
z > \zcut
\bigg(\frac{\theta_{12}}{R}\bigg)^\beta,\qquad\quad
z \equiv \frac{\text{min}(p_{t1},p_{t2})}{p_{t1}+p_{t2}},
\end{equation}
with $p_{ti}$ the transverse momentum of the subjet $j_i$,
$\theta_{12}$ their angular separation in the rapidity-azimuthal angle
plane and $R$ the jet radius.
If the symmetry condition Eq.~(\ref{eq:softdrop-condition}) is not
met, the procedure is recursively applied to the subjet with the
largest $p_t$.
Eq.~(\ref{eq:softdrop-condition}) with $\beta=0$ corresponds to the
mMDT,\footnote{Throughout this paper, we assume that the $\mu$
  parameter of the mMDT is set to 1. 
  Choosing a small value for $\mu$ would have an effect similar to
  that of a (recursive) $N$-subjettiness cut, as discussed
  in~\cite{Dasgupta:2015lxh}.}
while SD generalises it to the case of
$\beta\neq 0$.
Note that to some extent mMDT and SD have both tagging and grooming
properties.
When we use mMDT and SD together, the $\zcut$ parameter of SD will be
renamed $\zetacut$ in order to avoid confusion.

$N$-subjettiness is defined as follows: for a given jet, one finds a
set of $N$ axes $a_1,\dots,a_N$ (see below) and introduces
\begin{equation}\label{eq:tauN}
\tau_N^{(\beta_\tau)} = \frac{1}{p_t R^{\beta_\tau}} \sum_{i\in{\rm jet}}
  p_{ti} \min(\theta_{ia_1}^{\beta_\tau},\dots,\theta_{ia_N}^{\beta_\tau}),
\end{equation}
where the sum runs over all the constituents of the jet, of momentum
$p_{ti}$ and with an angular distance
$\theta_{ia_j}=\sqrt{\Delta y_{ia_j}^2+\Delta\phi_{ia_j}^2}$ to the
axis $a_j$;
$\beta_\tau$ is a free parameter and in what follows we will concentrate on
the case $\beta_\tau=2$. This specific choice has shown good performance in
Monte-Carlo numerical simulations and considerably simplifies the
physical discussions below.
However, the techniques introduced in this paper straightforwardly
apply to other values of $\beta_\tau$, including the frequent
experimental choice $\beta_\tau=1$, and we will comment on
this in Section~\ref{sec:comparison-others} and
Appendix~\ref{app:beta1}.

We still need to specify how to choose the $N$-subjettiness axes. In
practice, there are several methods that one can use. Common choices
include using exclusive $k_t$ axes or using ``minimal'' axes, \ie use
the set of axes that minimise the $\tau_N$. We will instead consider the
case of exclusive axes obtained by declustering the result of a
generalised-$k_t$ with $p=1/2$
\cite{Cacciari:2008gp,Cacciari:2011ma}. The motivation behind this
choice has been explained in~\cite{Dasgupta:2015lxh} and is related to
the fact that, since it preserves the ordering in mass, it produces
results very close to the much more complex minimal axes.\footnote{For
  a generic $\beta_\tau$, one could use the generalised-$k_t$ algorithm
  with $p=1/\beta_\tau$, using the winner-takes-all (WTA) recombination
  scheme~\cite{Larkoski:2014uqa} for $\beta_\tau\le 1$ to avoid
  inconvenient recoil effects.}

Since weak bosons radiate less than QCD jets, the ratio
$\tau_{21}=\tau_2/\tau_1$ is expected to be smaller for weak bosons
and one imposes a cut $\tau_{21}<\taucut$ as a radiation constraint to
distinguish weak bosons from the QCD background.

\subsection{A useful graphical representation}\label{sec:lund}

\begin{figure}
\begin{minipage}[b]{0.45\textwidth}
  \caption{Lund diagram representing the phasespace available for an
    emission from the jet initial parton at an angle $\theta$ and
    carrying a momentum fraction $z$. The diagram shows a
    given emission (the solid dot) as well as lines with the same
    momentum fraction, $k_t$ and mass scales.}\label{fig:lund-base}
\vspace*{2.0cm}
\end{minipage}
\hfill
\includegraphics[width=0.48\textwidth]{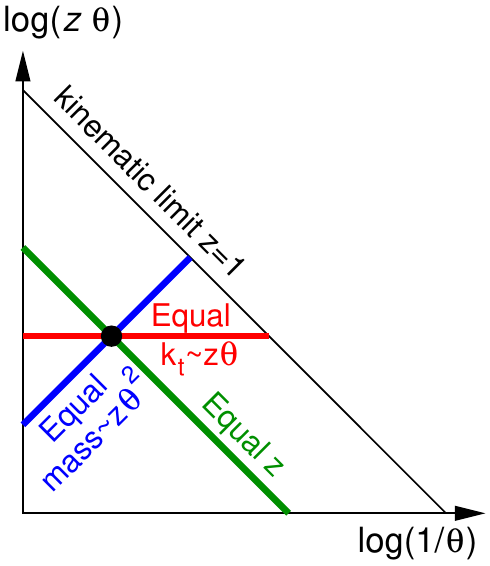}
\end{figure}

To guide our discussion, it is helpful to consider the
available phasespace for radiation inside a (QCD) jet in the
soft-and-collinear limit and see how the various methods under
consideration constrain that phasespace.
This is conveniently done using Lund diagrams~\cite{Andersson:1988gp}.
Consider an emission at an angle $\theta$ from the jet axis, carrying
a fraction $z$ of the transverse momentum of the parent parton.
Lund diagrams represent the two-dimensional phasespace for emissions
using the angle, or more precisely $\log(1/\theta)$, on the horizontal
axis, and the relative transverse momentum, $\log(k_t/p_{t,{\rm
    jet}})=\log(z\theta)$, on the vertical axis.
As shown in Fig.~\ref{fig:lund-base}, a line of constant momentum
fraction $z$ corresponds to a diagonal line with
$\log(k_t)=\text{constant}-\log(1/\theta)$ and a line of a given mass,
$m^2\sim z\theta^2$ in the soft and small-angle approximation,
corresponds to a diagonal line with
$\log(k_t)=\text{constant}+\log(1/\theta)$.

In the soft-and-collinear approximation, sufficient for the following
discussion, each emission comes with a weight
\begin{equation}
  d^2\omega = 
  \frac{2\alpha_s(k_t)C_R}{\pi}\, d\log(1/\theta)\,d\log(k_t)\label{eq:1}\,,
\end{equation}
with $C_R$ the colour factor of the parton initiating the jet, \ie
$C_F=4/3$ or $C_A=3$ respectively for quark and gluon jets.
The strong coupling constant, $\alpha_s$, is evaluated at a scale
equal to the transverse momentum of the emission relative to its
emitter.
Apart from running-coupling effects and subleading corrections in the
hard-collinear and soft-large-angle regions, this weight is uniform
over the Lund plane.

In the leading logarithmic approximation, the radiation in a jet is a
superposition of independent and strongly-ordered (primary) emissions
in that plane, as well as secondary emissions emitted from the primary
emissions and which can be represented as extra Lund triangles
(leaves) originating from each of the primary emissions, tertiary
emissions emitted from secondary ones, etc...
Leaves will be discussed in more detail below.

\begin{figure}
\centering{
\includegraphics[width=0.4\textwidth]{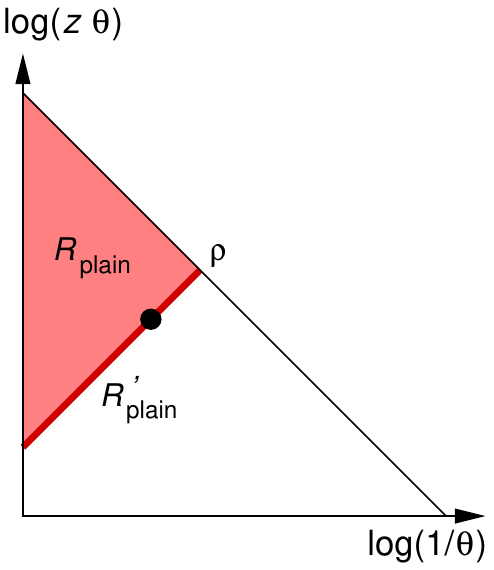}
\hfill
\includegraphics[width=0.4\textwidth]{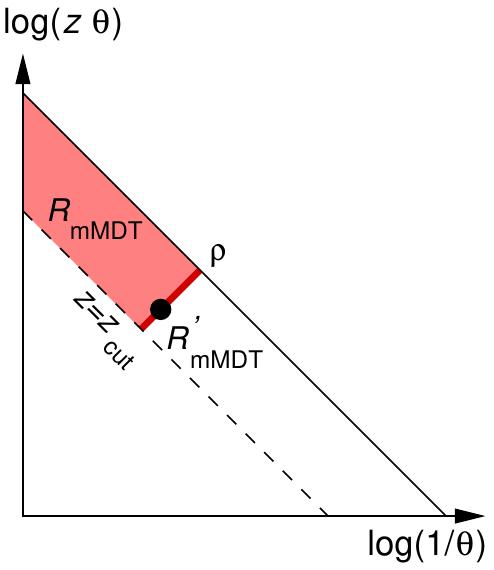}
}
\caption{Lund diagram representation for the phasespace regions
  relevant to the \fulljet jet mass (left) and the mMDT mass (right). The
  solid black point corresponds to the emission dominating the jet
  mass and can be anywhere along the solid red line. It gives the
  prefactor in the jet mass distribution. The shaded red area
  corresponds to the vetoed region yielding the Sudakov
  exponent.}\label{fig:lund-masses}
\end{figure}

To illustrate how one can use this pictorial representation to discuss
physics processes, let us consider the case of the (\fulljet) jet mass
distribution $m^2\!/\sigma\: d\sigma/dm^2$.
The corresponding Lund diagram is represented in the left panel of
Fig.~\ref{fig:lund-masses}.
One first needs an emission that provides the dominant contribution to
the mass of the jet, \ie an emission such that $m^2=z\theta^2p_t^2$
or such that $z\theta^2=\rho=(m/p_tR)^2$, where we have conveniently
normalised the angles in units of the jet radius $R$ and introduced the dimensionless
(squared) mass $\rho$ instead of $m^2$. For simplicity, we shall
assume a jet radius of $R=1$ from now on. 
The integrated weight for emissions that generate a (normalised) jet
mass equal to $\rho$ is\footnote{The $R$ use below in
  weights and Sudakov factors stands for ``Radiator'' and is not to
  be confused with the jet radius.}
\begin{equation}\label{eq:emission-rho}
R_{\text{\fulljet}}'(\rho)
 = \int d^2\omega\, \rho\delta(z\theta^2-\rho)
  \overset{{\text{f.c.}}}{=} \frac{\alpha_sC_R}{\pi}\log(1/\rho),
\end{equation}
where for the last equality we have illustrated the structure of the
answer in a fixed coupling
approximation, as indicated by the superscript ``f.c.''.
Modulo corrections induced by the running of the strong coupling, the
logarithmic behaviour basically comes from the integration over the
solid line of equal mass in the Lund representation.

We also need to impose that no emissions occur at larger mass. This
induces a Sudakov suppression $\exp[-R_{\text{\fulljet}}(\rho)]$
where\footnote{Technically, the exponential comes from the fact that,
  in the region $z\theta^2>\rho$, real emissions are vetoed while
  virtual contributions are present.}
\begin{equation}\label{eq:sudakov-rho}
R_{\text{\fulljet}}(\rho)
 = \int d^2\omega\, \Theta(z\theta^2>\rho)
  \overset{{\text{f.c.}}}{=} \frac{\alpha_sC_R}{2\pi}\log^2(1/\rho).
\end{equation}
The double-logarithmic behaviour corresponds to the shaded area
in the Lund diagram. Note that $R_{\text{\fulljet}}'(\rho)$ defined in
Eq.~(\ref{eq:emission-rho}) is the derivative of
$R_{\text{\fulljet}}(\rho)$ with respect to $\log(1/\rho)$.

In the end, the leading-logarithmic (LL) result for the cross-section
can be written as
\begin{equation}\label{eq:result-plainmass}
\left.\frac{\rho}{\sigma}\frac{d\sigma}{d\rho}\right|_{\text{\fulljet}}
  = R_{\text{\fulljet}}'(\rho)\,e^{-R_{\text{\fulljet}}(\rho)}.
\end{equation}
This expression has a simple graphical representation: a prefactor
corresponding to the emission setting the mass, the solid line in the
Lund diagram, and a Sudakov suppression for larger masses, the shaded
area in the Lund diagram.

Let us now consider the jet mass distribution after the application of
the mMDT.
This is represented in the right panel of Fig.~\ref{fig:lund-masses}.
In this case~\cite{Dasgupta:2013ihk}, emissions with $z<\zcut$ are
discarded by the mMDT recursive procedure,\footnote{Strictly speaking,
  only emissions with $z<\zcut$ and at an angle larger than the first
  emission with $z>\zcut$ will be discarded. This has no impact on the
  discussion of the jet mass since the difference only introduces a subleading
  correction. Subleading corrections to the mMDT/SD
  mass distribution are discussed in Ref.~\cite{Frye:2016aiz}.}
so that both the prefactor $R_{\text{mMDT}}'$ for having
an emission setting the jet mass and the Sudakov exponent
$R_{\text{mMDT}}$ are restricted to $z>\zcut$ and we have (assuming
$\rho\ll \zcut \ll 1$),
\begin{subequations}\label{eq:result-mmdtmass}
  \begin{align}
    \left.\frac{\rho}{\sigma}\frac{d\sigma}{d\rho}\right|_{\text{mMDT}}
    & = R_{\text{mMDT}}'(\rho)\,e^{-R_{\text{mMDT}}(\rho)},\\
    R_{\text{mMDT}}'(\rho)
    & = \int d^2\omega\, \rho\delta(z\theta^2-\rho) \Theta(z>\zcut)
      \overset{{\text{f.c.}}}{=} \frac{\alpha_sC_R}{\pi}\log(1/\zcut),\\
    R_{\text{mMDT}}(\rho)
    & = \int d^2\omega\, \Theta(z\theta^2>\rho) \Theta(z>\zcut)
      \overset{{\text{f.c.}}}{=} \frac{\alpha_sC_R}{\pi}[\log(\rho)\log(\zcut)-\frac{1}{2}\log^2(\zcut)].
  \end{align}
\end{subequations}
Compared to the \fulljet mass result, Eq.~(\ref{eq:result-plainmass}),
the prefactor is smaller but the Sudakov suppression is also less
important. In practice, we will therefore have a suppression of the
QCD background at intermediate masses but an increase at very small
masses.
More generally, we see that in order to have a large suppression of
the QCD background, we want a method that keeps the prefactor small
but gives a large Sudakov suppression.
This will be a key element of our dichroic approach.

Next, we consider the signal
(electroweak boson) jets in the context of Lund diagrams.
For, say, a $W$ boson, the original splitting, $W\to q\bar q$ occurs on
a line of constant mass $m=m_W$ and, since the corresponding splitting
function does not have a $1/z$ divergence at small $z$, this splitting
will be concentrated close to the large-$z$ end of that constant-mass
line, with the small-$z$ tail exponentially suppressed (in our
logarithmic choice of axes).
As a direct consequence, no emissions are possible at larger mass and
there will not be any Sudakov factor for that region.
For simplicity in our discussion below, we will assume a constant
splitting function in $z$, which would be the case \eg for a Higgs
boson or an unpolarised $W$ boson.
For an mMDT $\zcut$ condition, this yields a signal efficiency of
$1-2\zcut$. 
Subsequent emissions from the original $q\bar q$ pair will happen as if
they were secondary emissions from these two quarks, \ie essentially in two
separate Lund planes each of them restricted to angles smaller than the
separation $\theta_{q\bar q}$ between the two quarks, because of
angular ordering. 
One of those Lund planes (that for the softer of the $q\bar q$ pair)
will be represented as a leaf, cf.\ Fig.~\ref{fig:lund-nsubjettiness}.

Now that we have discussed how mass distribution and radiation
constraints are represented in terms of Lund diagrams, we will use
Lund diagrams to discuss more complex substructure methods, 
leaving corresponding analytic expressions to section~\ref{sec:analytic}.

\subsection{Radiation constraints ($N$-subjettiness)}\label{sec:nsubjettiness}

Let us now examine how a cut on $N$-subjettiness on the \fulljet jet
affects the pattern of allowed radiation.
Our discussion will be in a context where the \fulljet jet has a
specified mass, denoted through $\rho$.
The constraints imposed by a cut on the $N$-subjettiness ratio
$\tau_{21}$ can then again be presented quite straightforwardly in terms of
Lund diagrams, at least in the small $\tau_{21}$ limit, which is what
we will consider in our discussion.

\begin{figure}
\centering{
\includegraphics[width=0.4\textwidth]{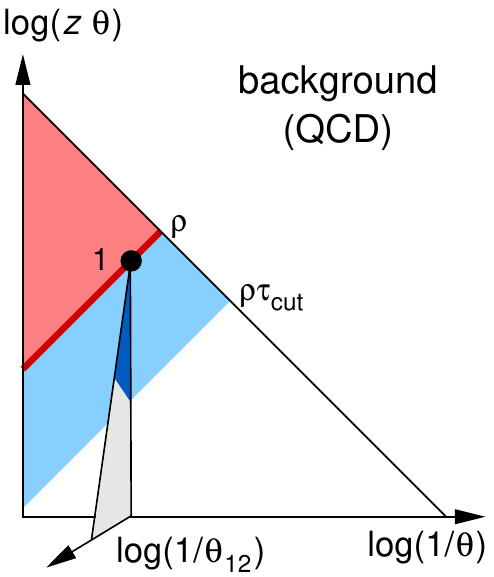}%
\hfill%
\includegraphics[width=0.4\textwidth]{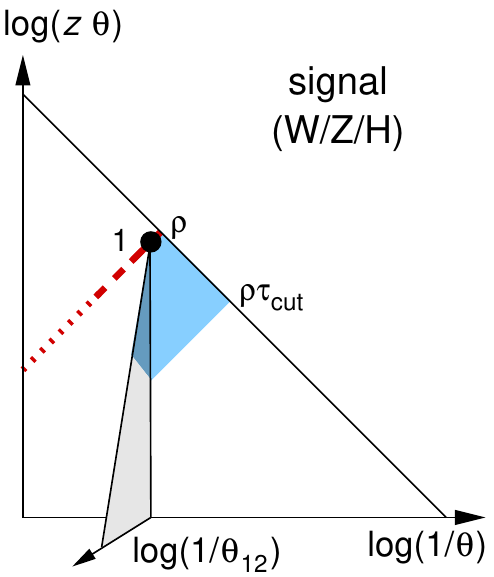}%
}
\caption{Lund diagram for QCD background jets (left) and signal jets (right)
  corresponding to the requirement of a given \fulljet jet mass with a
  cut on the $N$-subjettiness ratio $\tau_{21}$. The red shaded region
  (present only in the background case) corresponds to the Sudakov
  vetoed region for the mass, as in 
  Fig.~\ref{fig:lund-masses}, together with the prefactor for having
  an emission on the solid red line. 
  The blue shaded region
  corresponds to the additional veto coming from the cut on
  $N$-subjettiness.
  The dashed/dotted red line for the signal case represents the fact that,
  for signal jets,  small-$z$ configurations are
  exponentially suppressed.
  The region that emerges from the plane is referred to as a ``leaf''
  and in the left-hand diagram represents secondary emissions from
  emission $1$, while in the right-hand diagram it represents
  emissions from the softer of the two prongs of the decay.
}\label{fig:lund-nsubjettiness}
\end{figure}

Say that we have a first emission with an angle $\theta_1$ and momentum
fraction $z_1$ that dominates the jet mass, $\rho=z_1\theta_1^2$. 
It can be shown~\cite{Dasgupta:2015lxh} that, in our
leading-logarithmic approximation, $\tau_N$ (with $\beta_\tau=2$) will be
dominated by the $N^{\text{th}}$ largest $z\theta^2$.
We therefore have $\tau_1\equiv \rho$ and imposing a cut
$\tau_{21}<\taucut$ is equivalent to vetoing emissions down to a
``mass-like scale'' $z\theta^2=\rho\taucut$, for both primary and
secondary emissions.
This is represented in Fig.~\ref{fig:lund-nsubjettiness} for QCD and
signal jets, where the extra constraint on $N$-subjettiness
corresponds to an extra Sudakov factor represented by the blue shaded
region.
In the background case, the leaf that emerges from the plane
corresponds to a region of secondary emissions, while in the signal it
corresponds to the region of emissions from the softer of the $q\bar q$
pair. 
Assuming a background mainly consisting of quark jets, the main parts
of the plane in the two figures are both associated with a $C_F$
colour factor, while the leaf in the background case is associated
with a $C_A$ colour factor, in contrast with the $C_F$ factor for the
signal, and correspondingly represented with a darker shade of
blue.

We see that we now have a Sudakov suppression for both the signal and
the QCD background. Since the vetoed area is larger for the background
than for the signal, the former is more suppressed than the latter,
implying a gain in discriminating power.
Furthermore, since, for a given $\taucut$, the vetoed area increases
when $\rho$ gets smaller, the discriminating power will also be larger
for more boosted jets.

\section{Dichroic subjettiness ratios}\label{sec:new}

\subsection{Combining mMDT/SD with $N$-subjettiness}\label{sec:tagger+nsub}

We can now present the main proposal of this paper concerning the
dichroic combination of a tagger with a radiation constraint.
The discussion below assumes that we use SoftDrop or the
modified MassDrop tagger as our tagger and a cut on $\tau_{21}$ as a
radiation constraint, but we believe that the core argument can also
be applied to other shapes, for example to energy correlation
functions~\cite{Larkoski:2013eya,Larkoski:2014gra,Moult:2016cvt}.

Let us consider a high-$p_t$ large-radius ($R \simeq 1$) jet on which
we have applied an mMDT (or SD) tagger.
The original large-radius jet will be called the \fulljet jet.
The part of the jet that remains after the mMDT/SD tagging procedure
will be called the tagged jet, and has an angular size comparable to
the angle between the two hard prongs identified by the tagger.
The $N$-subjettiness variables $\tau_1$ and $\tau_2$ can be evaluated
either on the full or the tagged jet and there are three combinations
of interest:
\begin{subequations}
  \begin{alignat}{2}
  &\tauT&&\equiv \frac{\tau_2^\text{tagged}}{\tau_1^\text{tagged}}\label{eq:tauT},\\
  &\tauP&&\equiv \frac{\tau_2^\text{\fulljet}}{\tau_1^\text{\fulljet}}  \label{eq:tauP},\\
  &\tauD&&\equiv \frac{\tau_2^\text{\fulljet}}{\tau_1^\text{tagged}} \label{eq:tauD}.
  \end{alignat}
\end{subequations}
The first two options are currently widely used in the literature (see
\eg
\cite{Behr:2015oqq,Aad:2015typ,Khachatryan:2016cfa,Dolen:2016kst,Khachatryan:2016mdm}
for recent examples).
The third, ``dichroic'', option is a new combination, and is the
subject of this paper.\footnote{One can be tempted to also consider a
  fourth option where $\tau_1$ is computed on the \fulljet jet and
  $\tau_2$ on the tagged jet. It is straightforward to show, following
  the same arguments as below, that this is not the best combination,
  as one might expect intuitively.}

To understand how these different variants work, we will take two
approaches.
First we will consider what values of $\tau_{21}$ arise for different
kinematic configurations involving three particles in the jet, \ie
two emissions in the case of QCD jets, and the original two prongs plus
one additional emission in the case of signal jets.
Then we will use this information to understand how a cut on
$\tau_{21}$ constrains the radiation inside the jet.

During this discussion it will be useful to keep in mind the core
difference between signal and background jets.
In the case of the background jets, the whole Lund plane and the leaf
can contain emissions, as shown in Fig.~\ref{fig:lund-nsubjettiness}(left).
In the case of signal jets, emissions are mostly limited to the region
shown in blue in Fig.~\ref{fig:lund-nsubjettiness}(right), \ie at
angles smaller than the decay opening angle and transverse momenta
smaller than the mass.
The leaves in the two cases have different colour factors, however we
will neglect this aspect in our discussion.\footnote{At low $p_t$ a
  significant part of $\tau_{21}$'s discriminating power is arguably
  associated with the leaf and, for gluon-initiated background jets,
  with the part of the main Lund plane that is at small angles
  compared to the decay opening. 
  This is mostly equivalent to quark--gluon discrimination, which is
  known to be only moderately
  effective~\cite{Gallicchio:2011xq,Gallicchio:2012ez,Larkoski:2013eya,Badger:2016bpw}
  and not to improve significantly at high-$p_t$.
  These effects are included in the analytic calculations of
  section~\ref{sec:analytic}. 
  }
Rather we will concentrate on the differences that arise at large
angle, \ie from the different coherent radiation patterns of coloured
versus net colour-neutral objects.

\begin{figure}[t]
\centering
\begin{tabular}{l|C{1.85cm}C{1.85cm}|C{1.85cm}C{1.85cm}|C{1.85cm}C{1.85cm}|}
& \multicolumn{2}{c|}{Case 1}
& \multicolumn{2}{c|}{Case 2}
& \multicolumn{2}{c|}{Case 3}\\
\cline{2-7}
& \multicolumn{2}{c|}{$z_c\theta_c^2\ll z_b\theta_b^2\ll z_a\theta_a^2$}
& \multicolumn{2}{c|}{$z_b\theta_b^2\ll z_c\theta_c^2\ll z_a\theta_a^2$}
& \multicolumn{2}{c|}{$z_b\theta_b^2\ll z_a\theta_a^2\ll z_c\theta_c^2$}\\
\cline{2-7}
& \multicolumn{2}{c|}{\includegraphics[width=0.265\textwidth]{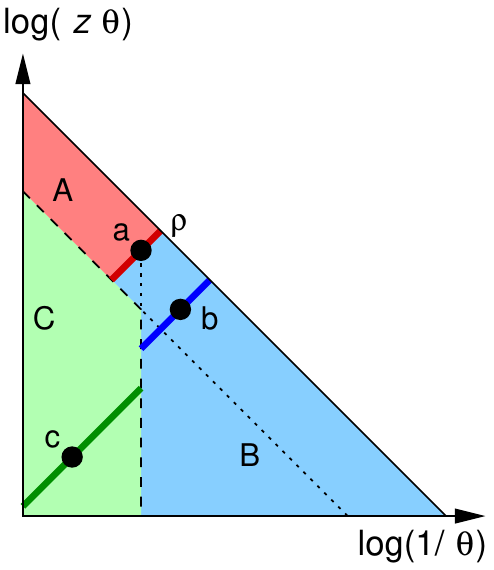}}
& \multicolumn{2}{c|}{\includegraphics[width=0.265\textwidth]{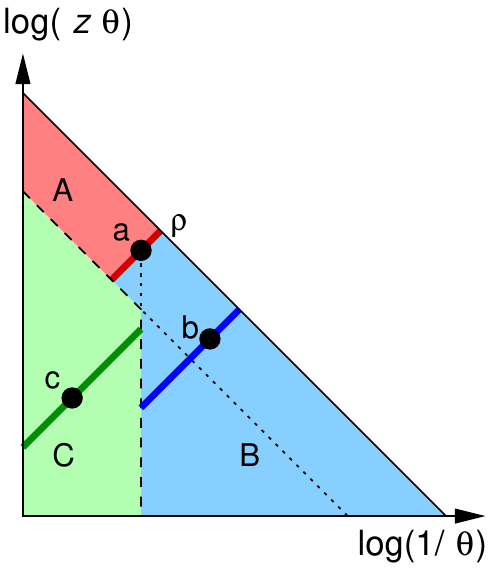}}
& \multicolumn{2}{c|}{\includegraphics[width=0.265\textwidth]{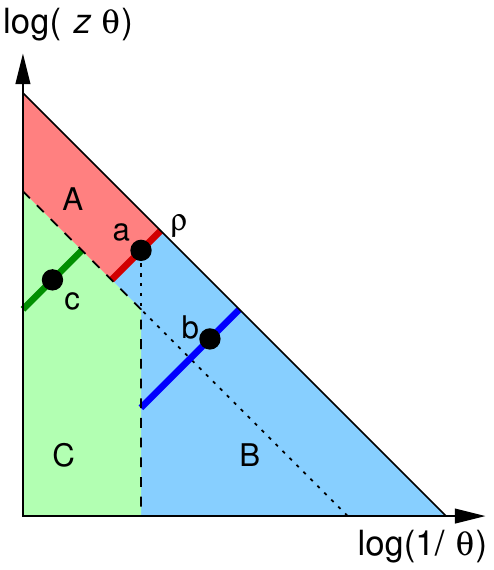}}\\
\cline{2-7}
& bkg & sig 
& bkg & sig 
& bkg & sig \\
\hline
$\tauT$
 & {\bf b/a} & b/a
 &      b/a  & b/a
 &      b/a  & b/a \\
\hline
$\tauP$
 & {\bf b/a} & b/a
 & {\bf c/a} & b/a
 &      a/c  & b/a \\
\hline
$\tauD$
 & {\bf b/a} & b/a
 & {\bf c/a} & b/a
 & {\bf a/a} & b/a  \\
\hline
\end{tabular}
\caption{Schematic representation of three possible kinematic
  configurations for the combination of $\tau_{21}$ with mMDT/SD
  (shown specifically for mMDT or SD with $\beta=0$).
  In each Lund diagram, emission ``a'' corresponds to the emission
  that dominates the mMDT/SD jet mass.
  This defines three regions: region A (red) is vetoed by mMDT, region
  B (blue) contains the constituents of the mMDT/SD jet and region C
  (blue) is the difference between the mMDT/SD jet and the \fulljet jet.
  Emissions ``b'' and ``c'' are respectively in regions B and C, and
  the three plots correspond to three different orderings of
  $z_c\theta_c^2$ compared to $z_a\theta_a^2$ and $z_b\theta_b^2$.
  The table below the plots shows the corresponding value of
  $\tau_{21}$ for both the QCD background (where all three regions
  have to be included) and the signal (where only regions A and B are
  present).
  For simplicity, ``b/a'' stands for
  $(z_b\theta_b^2)/(z_a\theta_a^2)$,
  and so forth.
  \label{fig:combination-3cases}}
\end{figure}

We consider the situation where, after the tagger has been applied,
the tagged jet mass is dominated by emission ``a'', \ie
$\rho\approx z_a\theta_a^2$ (in the case of the signal jet this is the
softer of the two prongs).
The Lund-plane phasespace can then be separated into 3 regions
depicted in Fig.~\ref{fig:combination-3cases}.
Region A (in red) is the region that is constrained to be free of
radiation by the fact that the tagger has triggered on emission a.
This corresponds to the region where both $z\theta^2>z_a \theta_a^2$
and Eq.~(\ref{eq:softdrop-condition}) are satisfied.
It is responsible for the Sudakov exponent associated with the
tagger, cf.\ Eq.~(\ref{eq:result-mmdtmass}).

Of the remaining phasespace, region B (blue) corresponds to emissions
that are contained inside the tagged jet.
It is populated in both signal and background cases.
It contains not only emissions that satisfy the mMDT/SD condition
($z>\zcut$ in the case of mMDT), but also emissions with
$z\theta^2<z_a \theta^2_a$ and $\theta < \theta_a$, due to the
Cambridge/Aachen declustering used by mMDT/SD.
Region C (green) corresponds to emissions that are in the original
\fulljet jet, but not in the tagged jet.
It is uniformly populated in the background case, while in the signal
case it is mostly empty of radiation, except at the left-hand edge
(initial-state radiation) and the right-hand edge (leakage of
radiation from the colour-singlet $q\bar q$ decay).
The emission with the largest $z\theta^2$ in each of regions B and C
will respectively be labelled $b$ and $c$ and we will assume strong
ordering between emissions, as in section~\ref{sec:setup}.

There are three kinematic cases to consider for the relative
$z\theta^2$ ordering of emissions $a$, $b$ and $c$, cf.\
Fig.~\ref{fig:combination-3cases}. 
In each case, Fig.~\ref{fig:combination-3cases} gives the result for
each of the $\tau_{21}$ variants, for both background and signal.
The signal case simply assumes that there are no emissions in region
C, which is appropriate in a double-logarithmic approximation.
The results are expressed as a shorthand, $i/j\equiv z_i\theta_i^2
/ z_j\theta_j^2$.

The case of the signal is particularly simple: since
$z_b\theta_b^2 < z_a\theta_a^2$ and there is nothing in region C,
all variants give $\tau_{21} = z_b\theta_b^2/z_a\theta_a^2$.
Given that the signal result is always the same, the performance of
the signal/background discrimination will be best for the method that
gives the largest background $\tau_{21}$ result (recall that one
enhances signal relative to background by requiring
$\tau_{21}<\taucut$).

Let us examine the background separately for each of the three kinematic
cases shown in Fig.~\ref{fig:combination-3cases}:
\begin{enumerate}[topsep=5pt,itemsep=0pt]
\item For $z_a\theta_a^2 \gg z_b\theta_b^2 \gg z_c\theta_c^2$, all
  three $\tau_{21}$ variants give the same result as for the signal,
  $z_b\theta_b^2/z_a\theta_a^2$.\footnote{Even if the signal and
    background have the same value, the different colour factor of the
    leaf, discussed earlier, still ensures discriminating power, because
    $z_b\theta_b^2/z_a\theta_a^2$ tends to be smaller for $C_F$ colour
    factors (signal) than for $C_A$ colour factors (background).
  }
\item For
  $z_a^2 \theta_a^2 \gg z_c^2 \theta_c^2 \gg z_b^2 \theta_b^2$,
  $\tauT$ is still still given by $z_b\theta_b^2/z_a\theta_a^2$, but
  $\tauP$ and $\tauD$ now both take the larger value of
  $z_c\theta_c^2/z_a\theta_a^2$.
  They should therefore perform better in this case.
\item Finally, for
  $z_c^2 \theta_c^2 \gg z_a^2 \theta_a^2 \gg z_b^2 \theta_b^2$,
  $\tauT$ is again given by $z_b\theta_b^2/z_a\theta_a^2$;
  $\tauP$ is given by $z_a\theta_a^2/z_c\theta_c^2$, since $\tau_1$ is
  dominated by emission $c$, while $\tau_2$ is dominated by emission
  $a$. 
  Depending on the exact configuration, $\tauP$ may be larger or
  smaller than $z_b\theta_b^2/z_a\theta_a^2$ and so may or may not be
  advantageous.
  $\tauD$ has a value of $z_a\theta_a^2 / z_a\theta_a^2 = 1$, which is
  always larger than the signal and larger than the other two
  variants.
\end{enumerate}
Overall therefore, $\tauD$ is expected to be the best of the three
variants.

\begin{figure}
\subfloat[][$\tauT$]{\label{fig:combination-3lund-i}%
  \includegraphics[width=0.3\textwidth]{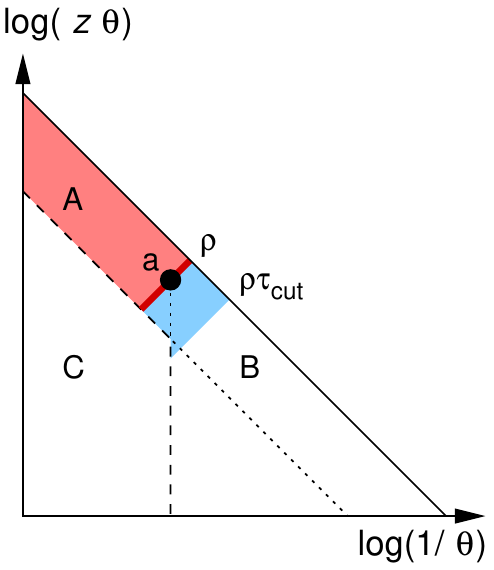}}\hfill
\subfloat[][$\tauP$]{\label{fig:combination-3lund-ii}%
  \includegraphics[width=0.3\textwidth]{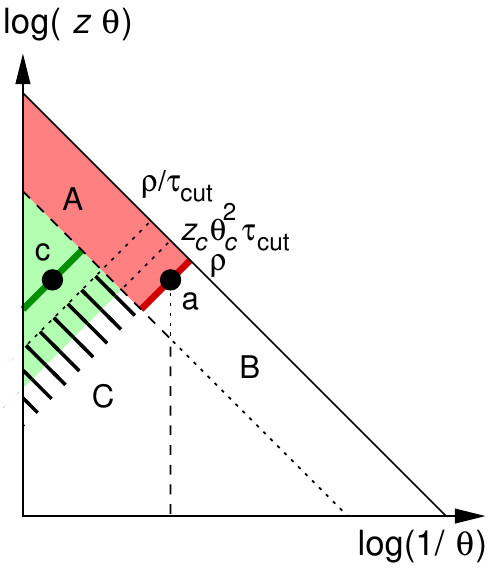}}\hfill
\subfloat[][$\tauD$ and $\tauP$]{\label{fig:combination-3lund-iii}%
  \includegraphics[width=0.3\textwidth]{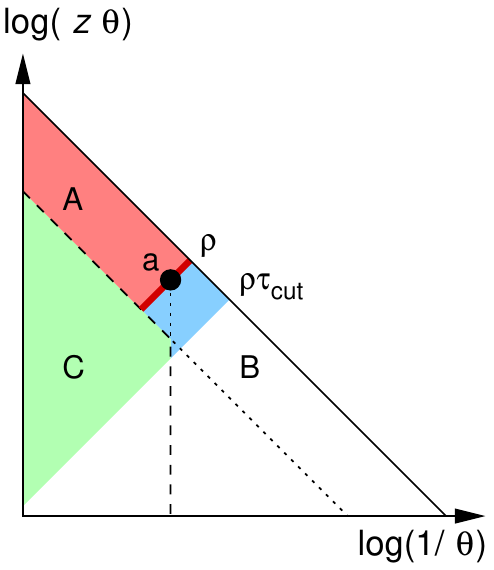}}
\caption{Regions where real emissions are vetoed when
  combining a mMDT/SD tagger with a cut on $\tau_{21}$.
  See text for details.
}\label{fig:combination-3lund}
\end{figure}

Alternatively, we can also see the benefit of the dichroic combination by
examining directly how emissions are constrained when one applies a
given cut on the $\tau_{21}$ ratio, similarly to the discussion in
Section~\ref{sec:nsubjettiness}.
We have represented the Lund diagrams relevant for our discussion in
Fig.~\ref{fig:combination-3lund}, where we have used the same regions
A, B and C as in the above discussion.

We start by considering a jet for which we already have applied the
mMDT/SD procedure, resulting in a (mMDT/SD) mass $\rho$ dominated by
emission ``a''.
This automatically comes with a mMDT/SD prefactor and Sudakov
suppression represented by the solid red line and shaded light red
area (region A) in Fig.~\ref{fig:combination-3lund}, guaranteeing that
there are no emissions at larger mass kept by the mMDT/SD.

For $\tauT$, emissions in region B are vetoed down to a mass scale
$\rho\taucut$ while emissions in region C, \ie outside the mMDT/SD
tagged jet, are left unconstrained.
This results in the (additional) Sudakov suppression given by the blue
area (region B) in Fig.~\ref{fig:combination-3lund}\subref{fig:combination-3lund-i}.

The situation for $\tauP$ is a bit more involved and we have three
cases to consider.
The first case is when there is (at least) one emission in region C
with $z\theta^2>z_a\theta_a^2/\taucut\equiv \rho/\taucut$ and is represented in
Fig.~\ref{fig:combination-3lund}\subref{fig:combination-3lund-ii}. Let us then call emission ``c''
the emission in region C with the largest $z\theta^2$, which thus
comes with a Sudakov suppression imposing that there are no other
emissions in region C with $z\theta^2>z_c\theta_c^2$. 
Emission ``c'' will dominate $\tau_1$ so that the cut on
$\tau_{21}$ will come with an extra suppression factor in region C
extending from $z_c\theta_c^2$ down to
$z\theta^2=z_c\theta_c^2\taucut$. 
Consequently, all emissions down to
$z_c\theta_c^2\taucut$ are vetoed as depicted in
Fig.~\ref{fig:combination-3lund}\subref{fig:combination-3lund-ii}.
The second case is when the emission in region C with the largest
$z\theta^2$ satisfies $z_a\theta_a^2\equiv
\rho<z_c\theta_c^2<\rho/\taucut$.
This region, represented by the hatched area in
Fig.~\ref{fig:combination-3lund}\subref{fig:combination-3lund-ii}, is entirely forbidden because it
would give a value of $\tau_{21} \ge z_a\theta_a^2/z_c\theta_c^2$
which is always larger than $\taucut$.
The third case is when there are no emissions in region C with
$z\theta^2>\rho$. This directly comes with a Sudakov suppression in
region C vetoing emission down to $z\theta^2=\rho$. In this case,
$\tau_1$ is dominated by emission ``a'' and the constraint on
$\tau_{21}$ further vetoes emissions with $\rho\taucut<z\theta^2<\rho$ in
both regions B and C. These two vetoes combine to vetoing all emission
down to $\rho\taucut$ as represented in
Fig.~\ref{fig:combination-3lund}\subref{fig:combination-3lund-iii}.

If instead we use our new $\tauD$ variable, we are always in the
situation of Fig.~\ref{fig:combination-3lund}\subref{fig:combination-3lund-iii}, where we veto all
emissions down to a mass scale $\rho\taucut$ in both regions B and C. 
This new version therefore comes with the strongest Sudakov
suppression, \ie of the three $\tau_{21}$ variables it is the one
that, for background jets, is least likely to have a small $\tau_{21}$
value.
Given that the three $\tau_{21}$ variants behave similarly to each
other for signal, the signal-to-background discrimination should be
improved for the dichroic variant.

With our dichroic method, we actually recover the same overall Sudakov
suppression as the one we had when measuring the \fulljet jet mass and
cutting on the \fulljet $N$-subjettiness (see
Section~\ref{sec:nsubjettiness} and
Fig.~\ref{fig:lund-nsubjettiness}(left)).
The gain of our new method (\ref{eq:tauD}) compared to this \fulljet
$N$-subjettiness case comes from the fact that the prefactor
associated with the jet mass is now subject to the constraint imposed
by the tagger.
If we take for example the case of the mMDT, this prefactor would be
largely suppressed for the background --- going from
$\sim\alpha_s\log(1/\rho)$ for \fulljet $N$-subjettiness to
$\sim\alpha_s\log(1/\zcut)$ for the dichroic method --- while the
signal would only be suppressed by a much smaller factor $\sim
1-2\zcut$.
Additionally, measuring the tagged jet mass instead of the \fulljet jet
mass significantly reduces ISR and non-perturbative effects which
would otherwise affect the resolution of the signal mass peak (see
also \cite{Dasgupta:2015yua,ourYsplitter}).

Finally, we note that the gain in performance is expected to increase
for larger boosts due to region C getting bigger
(double-logarithmically in $\rho$).

\subsection{Dichroic subjettiness with SoftDrop (pre-)grooming}
\label{sec:tagger+nsub+groomer}

Since $\tauD$ uses $\tau_2$ computed on the \fulljet jet, including all the
soft radiation at large angles, we can expect this observable to be
quite sensitive to poorly-controlled non-perturbative effects ---
hadronisation and the Underlying Event --- and to pileup.

The standard strategy to mitigate these effects is to kill two birds
with one stone and to use mMDT (or SD) both as a two-prong tagger
and as a groomer, and impose the $\tau_{21}$ constraint on the
result. 
This is equivalent to the $\tauT$ variant discussed
(Fig.~\ref{fig:combination-3lund}(a)), with the drawback and loss of
performance described in the previous Section.

We show here how we can achieve a background rejection that is
larger than for $\tauT$ and more robust with respect to
non-perturbative effects than $\tauD$.
Conceptually, the idea is that the tagger and groomer achieve two
different tasks: the tagger selects a two-prong structure in the jet,
imposing a rather hard constraint on the soft radiation in order to do
so, leading to a small $R'$ prefactor for the jet mass.
This is not quite what we want from a groomer, which should get rid of
the soft-and-large-angle radiation while retaining enough of the jet
substructure to have some discriminating power when using radiation
constraints.

\begin{figure}
\centering
\begin{minipage}[t]{0.4\textwidth}
\mbox{ }\\
\includegraphics[width=\textwidth]{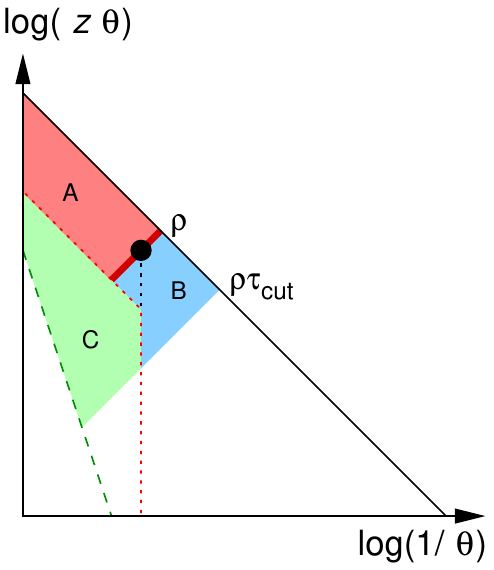}%
\end{minipage}
\hfill%
\begin{minipage}[t]{0.55\textwidth}
  \caption{Phasespace constraints on QCD jets obtained from our new
    combination including grooming: we first groom the jet, \eg
    with SoftDrop (SD). We then compute both the jet mass and $\tau_1$
    on the tagged jet (here using the mMDT), yielding the
    solid red line prefactor and the shaded red region (A) for the
    Sudakov exponent. 
    We then impose a cut on the $\tau_{21}$ ratio with $\tau_2$
    computed on the SD jet, leading to the extra shaded blue and green
    regions (B and C) for the Sudakov
    exponent.}\label{fig:lund-new-groomed}
\end{minipage}
\end{figure}

This suggests the following picture: we first apply a ``gentle''
grooming procedure to the jet, like a SoftDrop procedure with a
positive value of $\beta$. This is meant to clean the jet of the
unwanted soft junk\footnote{In the presence of pileup, one should
  still apply a pileup subtraction procedure~\cite{Altheimer:2013yza},
  like area--median
  subtraction~\cite{Cacciari:2007fd,Cacciari:2008gn},
  charged-track-based
  techniques~\cite{ATL-Pileup-2,Cacciari:2014jta,Krohn:2013lba}, the
  constituent subtractor~\cite{Berta:2014eza},
  SoftKiller~\cite{Cacciari:2014gra} or
  PUPPI~\cite{Bertolini:2014bba}.
  This can be done straightforwardly with SoftDrop and mMDT.} 
while retaining as much as possible the information about the
perturbative radiation in the jet.
We can then carry on with the dichroic method presented in the
previous Section, \ie use a more aggressive tagger, like
mMDT,\footnote{Or SD with a smaller value of $\beta$ than used in the
  grooming.} 
to compute the jet mass and $\tau_1$ and compute $\tau_2$ on the SD
(pre-)groomed jet:
\begin{equation}\label{eq:tauD-groomed}
\tauDG
   = \frac{\tau_2(\text{SD jet})}
          {\tau_1(\text{mMDT jet})}.
\end{equation}
This is depicted in Fig.~\ref{fig:lund-new-groomed}, where regions A
and B are the same as in the previous Section, but now region C
indicates the region where emissions are kept by the groomer but
rejected by the tagger.
Similarly, we can introduce
\begin{equation}\label{eq:tauP-groomed}
\tauPG
   = \frac{\tau_2(\text{SD jet})}
          {\tau_1(\text{SD jet})}.
\end{equation}
Note that we will always choose our mMDT-tagging and SD-grooming parameters
such that the tagged jet is the same whether tagging is performed before
or after grooming.
For mMDT-tagging with parameter $\zcut$ and SD-grooming 
with parameters $\zetacut$ and $\beta$, this implies
$\zetacut \le \zcut$ and $\beta \ge 0$.

Using the same arguments as in Section~\ref{sec:tagger+nsub}, we can
show straightforwardly that this method will have a larger rejection
than with the other two variants where one would be computing the jet
mass on the mMDT-tagged jet and the $\tau_{21}$ ratio either on the mMDT-tagged jet,
$\tauTG\equiv\tauT$, or on the SD-groomed jet, $\tauPG$, owing to a larger
Sudakov suppression of the background, for a similar signal efficiency.

Compared to the other possible situation where both the jet mass and
the $\tau_{21}$ ratio are computed on the SD-groomed jet, the
dichroic variant would have a smaller $R'$ prefactor, associated with
mMDT instead of SD. This again leads to a larger background
rejection.

Because of the initial grooming step, the groomed dichroic
subjettiness ration is expected to be less discriminating than the
ungroomed version introduced in Section~\ref{sec:tagger+nsub}.
Indeed, the associated Sudakov exponent is smaller since we have
amputated part of the soft-large-angle region. One should however
expect that this groomed variant will be less sensitive to
non-perturbative effects. 
Overall, there is therefore a trade-off between effectiveness, in
terms of achieving the largest suppression of the QCD background for a
given signal efficiency, and perturbative robustness, in terms of
limiting the sensitivity to poorly-controlled non-perturbative
effects.

\section{Performance in Monte-Carlo simulations}\label{sec:mc-study}

Let us now investigate the effectiveness and robustness of dichroic
subjettiness ratios in Monte-Carlo simulations, using
Pythia~8.186~\cite{pythia}, at a centre-of-mass energy of
$\sqrt{s}=13$~TeV.
Our signal sample consists of $WW$ events, while for the background we
use dijet events.
Jets are reconstructed with the anti-$k_t$~\cite{Cacciari:2008gp}
algorithm with $R=1$ and in determining signal and background
efficiencies we keep all jets above a given $p_t$ cut.\footnote{All
  jets in the signal sample above that cut are considered to be
  signal-like, even if they came from initial-state radiation; however
  such initial-state jets will have been relatively rare in our sample
  and so should not affect our final conclusions.}
We use the modified MassDrop tagger with $\zcut=0.1$ for the
2-prong tagging and vary the cut on the $\tau_{21}$ ratio. 
Whenever a SoftDrop (SD) grooming procedure is included, we use
$\zetacut=0.05$ and $\beta=2$ as illustrative parameter choices
(recall that
the SoftDrop condition is imposed as $z>\zetacut (\theta_{12}/R)^\beta$
instead of Eq.~(\ref{eq:softdrop-condition}), i.e.\ we use separate
symbols $\zcut$ and $\zetacut$ respectively for the parameters of mMDT
and SD).
Jet reconstruction and manipulation are performed with
FastJet~3.2.0~\cite{Cacciari:2005hq,Cacciari:2011ma} and
  fjcontrib~1.024~\cite{fjcontrib}.

\subsection{$N$-subjettiness and mass distributions with various $\tau_{21}$ ratios ($\beta_\tau=2$)}

\begin{figure}
  \begin{minipage}[c]{0.48\textwidth}
    \caption{%
      $\tau_{21}$ distributions for jets in dijet (solid lines) and
      $WW$ (dashed lines) events again imposing $p_t>2$~TeV and
      including SoftDrop grooming.
      Different colours correspond to different combinations of
      jets used for the computation of the jet mass, $\tau_1$ and
      $\tau_2$ as indicated in the legend, our new dichroic combination
      being plotted in black.
      We have selected jets with a mass is between 60 and 100~GeV.
      The cross-section used for normalisation, $\sigma$, is defined
      after the jet $p_t$ and mass cut, so that all curves integrate
      to one.
    }\label{fig:distribs-tau21}
  \end{minipage}\hfill
  \begin{minipage}[c]{0.48\textwidth}
    \centerline{\includegraphics[width=\textwidth,page=2]{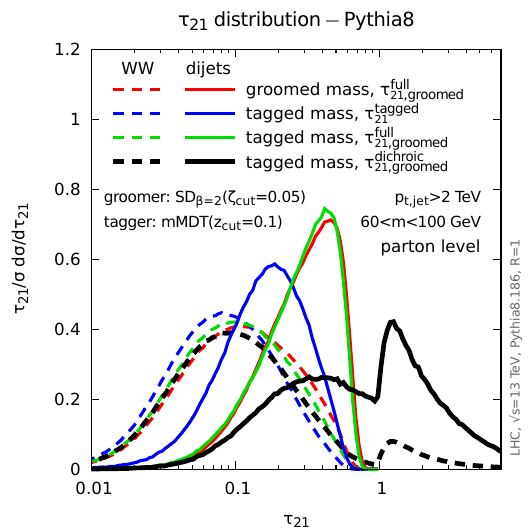}}
  \end{minipage}
\end{figure}

We start by examining the $\tau_{21}$ distribution. 
This is plotted in Fig.~\ref{fig:distribs-tau21} for both QCD jets
(solid lines) in dijet events and $W$ jets (dashed lines) in $WW$
events.
We select jets above 2~TeV and always apply SoftDrop grooming. 
In practice, we use parton-level events, and impose a cut on the
reconstructed jet mass (SD-groomed or mMDT-tagged) $60<m<100$~GeV.
We consider four cases: the $\tauPG$ distribution when we cut on the
SD-groomed mass and the $\tauT$, $\tauPG$ and $\tauDG$ distributions when
we cut on the mMDT-tagged mass. 
As expected, the distributions for signal ($W$) jets are
peaked at smaller values of $\tau_{21}$ than the corresponding
distribution for background (QCD) jets.
Fig.~\ref{fig:distribs-tau21} shows that all the signal distributions,
and in particular the three options where one measures the mMDT-tagged jet mass,
are very similar. This is in agreement with our discussion in the
previous Section.
Comparatively the background distributions look rather different. The
case where everything is computed from the mMDT-tagged jet (the solid blue
curve) peaks at smaller values of $\tau_{21}$ as expected from its
smaller Sudakov suppression, related to the fact that this combination
puts no constraints on large-angle emissions (region C in the previous
Section).
Furthermore, the dichroic combination, the solid black curve in
Fig.~\ref{fig:distribs-tau21}, is expected to have the largest
suppression and is indeed peaked at larger $\tau_{21}$ values,
translating into a larger discrimination against signal jets.

Note that the $\tau_{21}$ distribution for the dichroic combination also
shows a peak for $\tau_{12}>1$ that we have not discussed in our
earlier argumentation. 
This comes from events with multiple emissions in region C and will be
discussed briefly in our analytic calculations in
Section~\ref{sec:analytic}.

\begin{figure}
  \begin{minipage}[c]{0.48\textwidth}
    \centerline{\includegraphics[width=\textwidth,page=2]{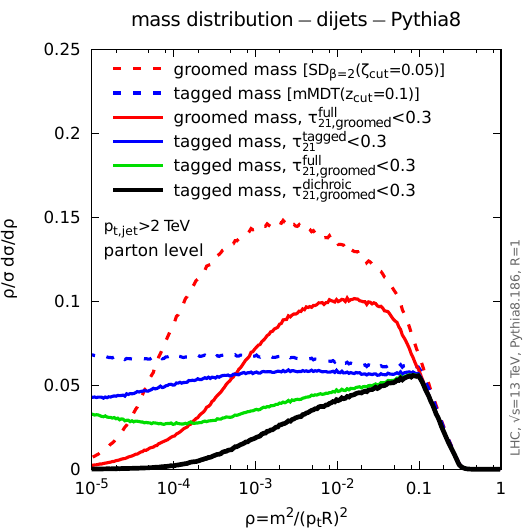}}
  \end{minipage}\hfill
  \begin{minipage}[c]{0.48\textwidth}
    \caption{%
      Mass distribution for QCD jets with $p_t>2$~TeV (anti-$k_t$,
      $R=1$) at parton level, including SoftDrop grooming.
      The dashed lines, in red for the SD-groomed jet and
      in blue for the mMDT-tagged jet, are the mass distributions 
      with no constraint on $N$-subjettiness.
      The solid lines have an additional cut $\tau_{21}<0.3$ with
      different combinations of jets used for the computation of the
      jet mass, $\tau_1$ and $\tau_2$ as indicated in the legend, our
      dichroic combination being plotted using a solid black line.
      The cross section used for normalisation, $\sigma$ is that for
      jets above the $p_t$ cut.
    }\label{fig:distribs-mass}
  \end{minipage}
\end{figure}

Results for the mass distribution obtained for background (QCD dijets)
jets at parton level (without UE) are presented in
Fig.~\ref{fig:distribs-mass}. 
As in Fig.~\ref{fig:distribs-tau21}, SoftDrop grooming has always been
applied prior to any additional tagging or $N$-subjettiness cut.
Again, we can identify most of the features discussed in
section~\ref{sec:new}.
First of all, if we compare the mMDT-tagged mass (dashed blue curve)
to the SD-groomed jet mass (dashed red curve) we see that the latter
is smaller than the former at small masses, owing to the larger
Sudakov factor $R_{\text{SD}}>R_{\text{mMDT}}$, but larger at
intermediate masses, due to the larger prefactor
$R_{\text{SD}}'>R_{\text{mMDT}}'$.

Then, we can consider the effect of the additional constraint on the
$\tau_{21}$ ratio, taken here as $\tau_{21}<0.3$ for illustrative
purpose.
If we compute $\tau_{21}$ on the same jet as for the mass ($\tauPG$ in
solid red
and $\tauT$ in solid blue for the SD-groomed and mMDT-tagged jets respectively),
we see that the cut reduces the background, that the reduction
increases for smaller masses and that the reduction is larger for the
SD-groomed jet than for the mMDT-tagged jet.
This last point is a reflection of the fact,
that the Sudakov suppression associated with the $N$-subjettiness cut
is larger when both the mass and $\tau_{21}$ are computed on the SD-groomed
jet (Fig.~\ref{fig:lund-nsubjettiness}(left)) than when both the mass
and $\tau_{21}$ are computed on the mMDT-tagged jet
(Fig.~\ref{fig:combination-3lund}(left)).
Then, when measuring the mMDT-tagged jet mass, one sees that computing
$\tau_{21}$ on the SD-groomed jet ($\tauPG$, the solid green curve in
Fig.~\ref{fig:distribs-mass}) shows a larger suppression than
computing $\tau_{21}$ on the mMDT-tagged jet, although the difference is
reduced at very small masses.
Finally, if we consider our new, dichroic case,
Eq.~(\ref{eq:tauD-groomed}) ($\tauDG$, the solid black curve), we see
a larger suppression than in all other cases, as expected from our
earlier arguments.

\subsection{Signal v.\ background discrimination and other performance measures}

\begin{figure}[p]
\centerline{
\includegraphics[width=0.48\textwidth,page=5]{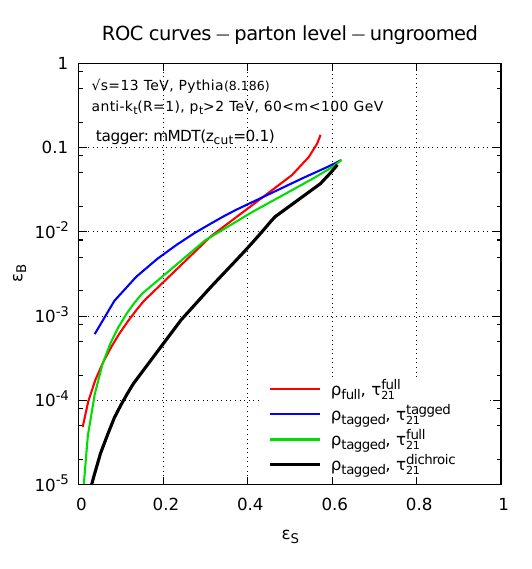}%
\hfill
\includegraphics[width=0.48\textwidth,page=7]{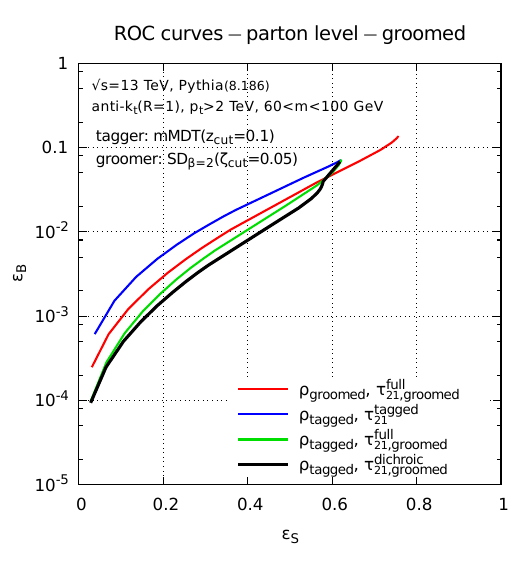}%
}
\caption{ROC curves for various $\tau_{21}$ combinations, \ie 
  background versus signal efficiency, at parton level.
  The left
  plot is obtained starting from the \fulljet jet, while for the right
  plot, a SoftDrop grooming step has been applied.
  The ROC curves are obtained by varying the cut on the $\tau_{21}$
  ratio.
  In all cases, we considered anti-$k_t$($R=1$) jets with
  $p_t>2$~TeV.}\label{fig:roc-parton}
\end{figure}

\begin{figure}
\centerline{
\includegraphics[width=0.48\textwidth,page=6]{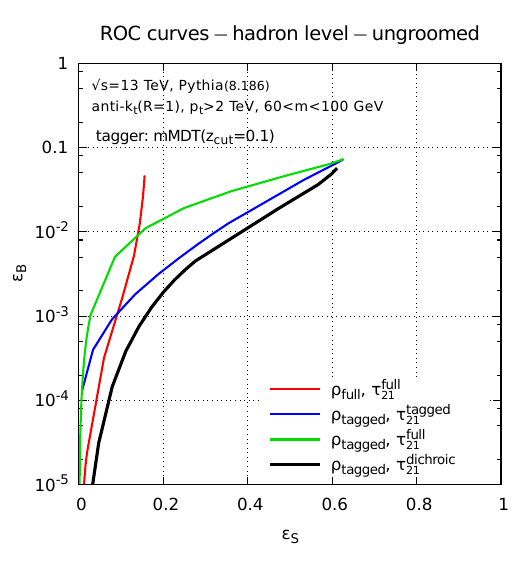}%
\hfill
\includegraphics[width=0.48\textwidth,page=8]{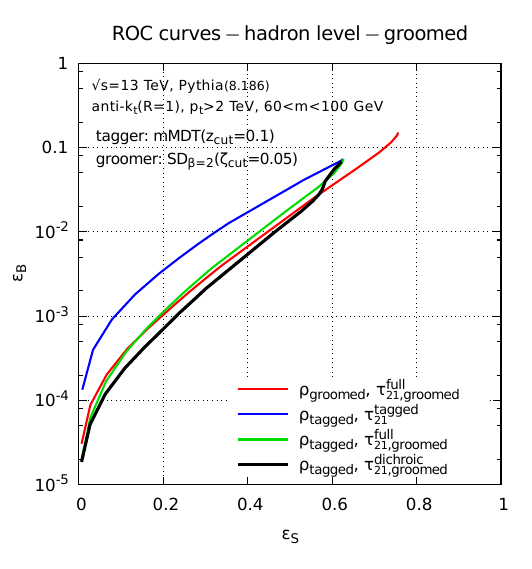}%
}
\caption{Same as figure as \ref{fig:roc-parton}, now for hadron level
  (including the Underlying Event).}\label{fig:roc-full}
\end{figure}

To further test the performance of our new method, we have also studied
ROC (receiver operating characteristic) curves, shown in
Fig.~\ref{fig:roc-parton} for parton-level 
simulations and in Fig.~\ref{fig:roc-full} for hadron-level events
including hadronisation and the Underlying Event.
In all cases, we impose the constraint that the mass is between 60 and
100~GeV.
Efficiencies are given relative to the inclusive cross-section for
having jets above our $p_t$ cut.

Let us first discuss the result of parton-level simulations,
Fig.~\ref{fig:roc-parton}, where the dichroic ratio is again
represented by the black curves.
Without grooming (the left-hand plot in the figure), 
our method shows a substantial improvement compared to all other
combinations considered, outperforming them by almost 30\% in
background rejection at a signal efficiency of 50\% and by more than a
factor of 2 at a signal efficiency of 40\%.
After SoftDrop grooming (right-hand plot), the dichroic method,
\ie computing the jet mass and $\tau_1$ on the mMDT-tagged jet and $\tau_2$
on the SD-groomed jet, still shows an improvement, albeit less impressive than
what is observed using the \fulljet jet to compute $\tau_2$.

If instead we consider the results at hadron level, including both the
perturbative parton shower as well as non-perturbative effects,
Fig.~\ref{fig:roc-full}, we see that the dichroic subjettiness ratio
still does a better job than the other variants but the gain is
smaller.
For example, measuring the mMDT-tagged mass with a cut on the groomed
dichroic ratio, $\tauDG$, the optimal choice in
Fig.~\ref{fig:roc-full}, is only slightly better than the next best
choice where one measures the SD-groomed mass and imposes a constraint on
$\tauPG$.
This is because in going from parton to hadron level, the
$\rho_\text{groomed}$--$\tauPG$ curve has moved down more than the
$\rho_\text{tagged}$--$\tauDG$ curve, \ie the former is getting a
significantly larger boost in its discriminating power from
non-perturbative effects.\footnote{%
  That there should be larger non-perturbative effects in the
  $\rho_\text{groomed}$--$\tauPG$ can be understood as follows:
  because $\rho_\text{groomed}$ accepts a larger fraction of signal
  events in a given mass window than $\rho_\text{tagged}$, to reach
  the same final efficiency the $\tau_{21}$ cut must be pushed closer
  to the non-perturbative region.
}
This is potentially problematic, because one does not
necessarily want signal-to-background discrimination power for a
multi-TeV object to be substantially driven by the physics that takes
place at a scale of $1\GeV$, physics that cannot, with today's
techniques, be predicted from first principles.
Additionally, phenomena happening on a scale of
$1\GeV$ are difficult to measure reliably.

It would be interesting to investigate non-perturbative effects in
greater depth, both analytically, \eg following the approach used
in~\cite{Dasgupta:2013ihk}, or by studying their dependence across
different Monte-Carlo generators and associated tunes.
However, for the purpose of this article, we limit ourselves to using
the results from Pythia~8.
In evaluating the overall performance of different $\tau_{21}$
combinations we will consider both the signal significance and the
size of non-perturbative effects.
We will use the following alternative to ROC curves.
For a given method and $p_t$ cut, we first determine the $\tau_{21}$
cut required to obtain a desired signal efficiency (at hadron
level). For that value of the $\tau_{21}$ cut, we can compute the
signal significance, defined as $\epsilon_S/\sqrt{\epsilon_B}$
(computed at hadron level) which is a measure of the discriminating
power of the method; we then estimate non-perturbative effects as the
ratio between the background efficiency at hadron level divided by the
background efficiency at parton level, which is a measure of
robustness against non-perturbative effects.
We will show results for a range of different signal-efficiency
choices and jet $p_t$ cuts.

\begin{figure}[t]
  \centerline{%
    \includegraphics[width=0.48\textwidth,page=2]{figs/np-effects-v-epsS.pdf}%
    \hfill%
    \includegraphics[width=0.48\textwidth,page=1]{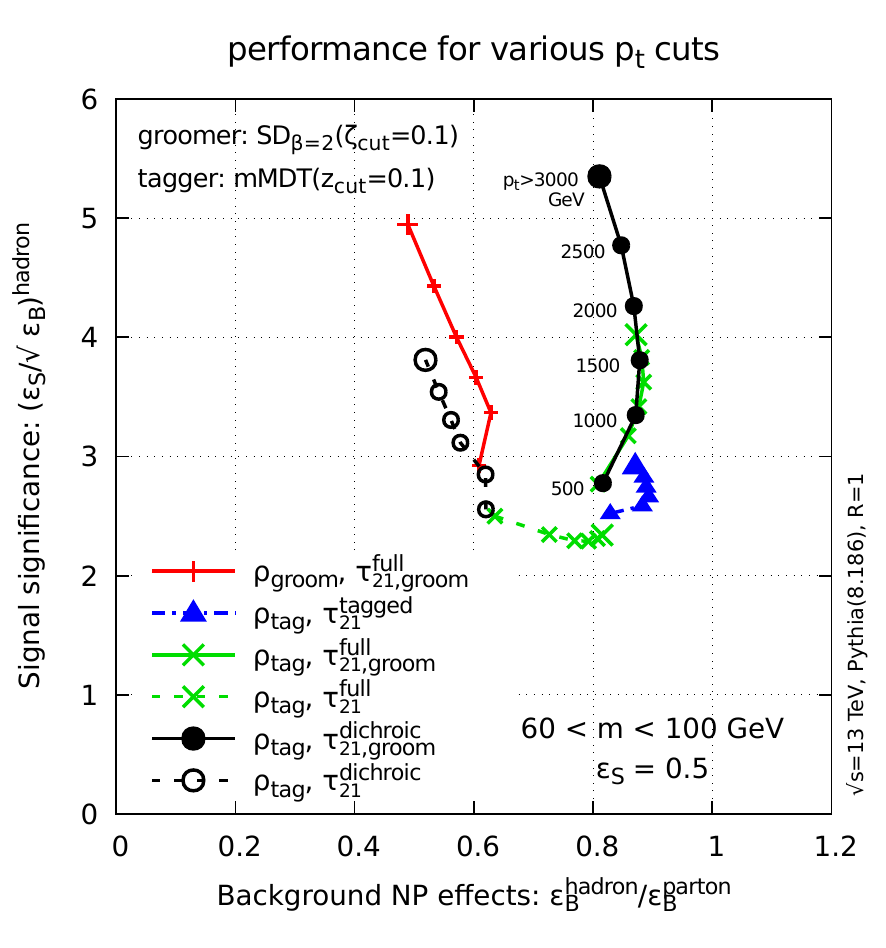}}
  \caption{Signal significance plotted versus the non-perturbative
    effects for the QCD background (defined as the ratio between the
    background ``fake'' tagging rate at hadron and parton level).
    Different curves correspond to different combinations indicated in
    the legend. For the solid curves, a SoftDrop ($\beta=2$ and
    $\zetacut=0.05$) grooming is applied, while no grooming
    is applied for the dashed curves.
    In the left-hand plot, we impose a 2~TeV $p_t$ cut on the initial
    jet. The symbols on each curve then correspond to a signal efficiency
    (computed at hadron level) ranging from 0.05 upwards in steps of 0.05, with the
    large symbol on each line corresponding to $\varepsilon_S=0.5$ and
    the efficiency at the right-hand extremity explicitly labelled.
    In the right-hand plot, the signal efficiency (computed at
    hadron level) is fixed to be 0.5 and the $p_t$ cut on the jet is varied between
    500~GeV and 3~TeV (in steps of 500~GeV, labelled explicitly for
    the groomed dichroic ratio), with the
    large symbol on each line corresponding to a 3~TeV cut.
  }\label{fig:np-effects}
\end{figure}

In Fig.~\ref{fig:np-effects}, which highlights the key performance
features of the dichroic method, we plot the signal
significance versus the non-perturbative effects for different
methods.
In the left-hand panel, the curves correspond to a range of $\tau_{21}$
cuts for jets with $p_t>2\TeV$. 
The points on the curves correspond to different signal efficiencies
(starting from 0.05, in steps of 0.05, and with $\epsilon_S=0.5$
indicated by a bigger point).
In the right-hand panel, the points on the curves correspond to
different $p_t$ cuts, with the $\tau_{21}$ cut adjusted (as a
function of $p_t$) so as to ensure a constant signal efficiency of
$0.5$.
To avoid the proliferation of curves, the result for the ungroomed
$\rho_\text{full}$--$\tauP$ is not shown since it is obvious from the
ROC curves in Figs.~\ref{fig:roc-parton} and \ref{fig:roc-full}(left)
that it is extremely sensitive to non-perturbative effects.

In both plots, we see that the dichroic method comes with larger
discriminating power with a relatively limited sensitivity to
non-perturbative effects, provided one first applies a grooming
step. 
Without the grooming step, one observes a much larger sensitivity to
non-perturbative effects, as one might expect.\footnote{It can also be
  shown that grooming largely reduces the impact of initial-state
  radiation as well (see also~\cite{Dasgupta:2015lxh}).}
It also appears that the performance gain increases when the boost,
\ie the jet $p_t$, increases. This was also expected from our
arguments in Section~\ref{sec:new}.
Finally, compared to the common setups in the literature, namely with
modified MassDrop tagging with a cut on $\tau_{21}$ applied either on
the mMDT ($\rho_\text{tag}$--$\tauT$, the dot-dashed blue curve) 
or on the \fulljet jet ($\rho_\text{tag}$--$\tauP$, the dashed green
curve), our dichroic method with grooming (solid black) gives up
to a factor of two improvement in signal significance, with comparable
non-perturbative effects.
Considering other combinations that have not been widely used
experimentally, $\tauPG$ with either a groomed ($\rho_\text{groom}$,
solid red) or a tagged ($\rho_\text{tag}$), solid green) jet mass both
perform well, however $\tauDG$ still remains the best, with an optimal
significance that is about $25\%$ larger, and smaller non-perturbative
corrections for any given signal significance.

\begin{figure}
  \centerline{%
    \includegraphics[width=0.48\textwidth,page=3]{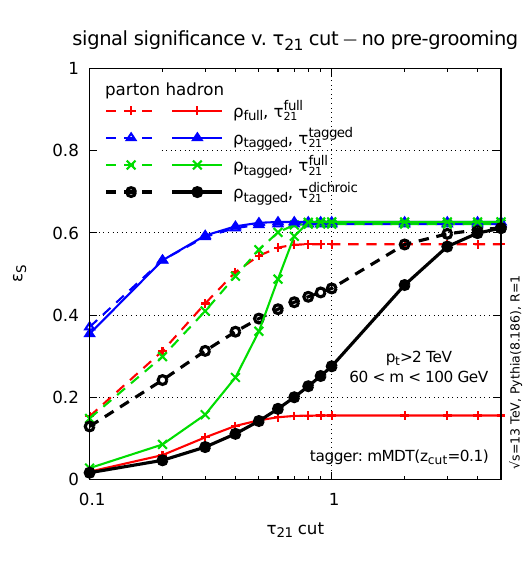}%
    \hfill%
    \includegraphics[width=0.48\textwidth,page=4]{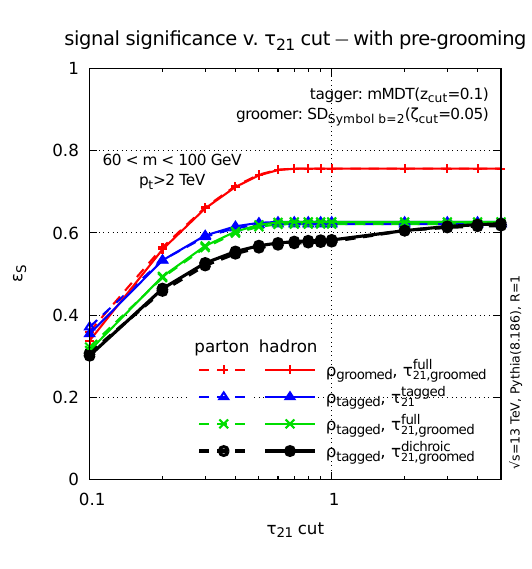}}
  \caption{Signal efficiency plotted as a function of the
    cut $\taucut$ on $\tau_{21}$ for all the combinations considered in
    Figs.~\ref{fig:roc-parton} and \ref{fig:roc-full}. 
    Solid curves correspond to hadron-level results while dashed
    curves are obtained at parton level. 
    The left plot is obtained starting from the \fulljet jet, while for
    the right plot, a SoftDrop grooming has been
    applied.
  }\label{fig:tau-v-eff}
\end{figure}

As a final check, we have studied the dependence of the signal
efficiency on the $\tau_{21}$ cut, as shown in
Fig.~\ref{fig:tau-v-eff}.
Comparing the left and right-hand plots, it appears clearly that applying
SoftDrop grooming helps to reduce non-perturbative
effects which otherwise significantly lower the signal efficiency.
It is also interesting to notice that without grooming, the signal
efficiency obtained with our dichroic method (the dashed black curve
on the left plot of Fig.~\ref{fig:tau-v-eff}) only reaches its plateau
for cuts on $\tau_{21}$ larger than 1 already at parton level.
This can likely be attributed to initial-state radiation in the jet at
angles larger than the decay angle of the $W$ boson. These effects are
strongly reduced by SoftDrop grooming (see also the discussion in
Section~\ref{sec:analytic}).

In the end, a more complete study would include variations of the SD
parameters and of the cuts on the mass. A brief investigation of
the SD parameters shows that our choice of $\beta=2$ and
$\zetacut=0.05$ seems a decent default, at least for the process and
kinematic domain under study.
However, in view of the good signal efficiency reached when computing
both the jet mass and $\tau_{21}$ with SoftDrop, it might also be
interesting to investigate our dichroic combination where we also use
SoftDrop for the tagger instead of the mMDT.
An extensive analytic study foreseen in a follow-up
paper~\cite{tau21-calculation} would allow for a systematic study of
these effects.
Such an analytic understanding could also be of use in the context of
building decorrelated taggers~\cite{Dolen:2016kst}.

\subsection{Brief comparison with other tools}\label{sec:comparison-others}

\begin{figure}[p]
\centerline{%
\includegraphics[width=0.49\textwidth,page=1]{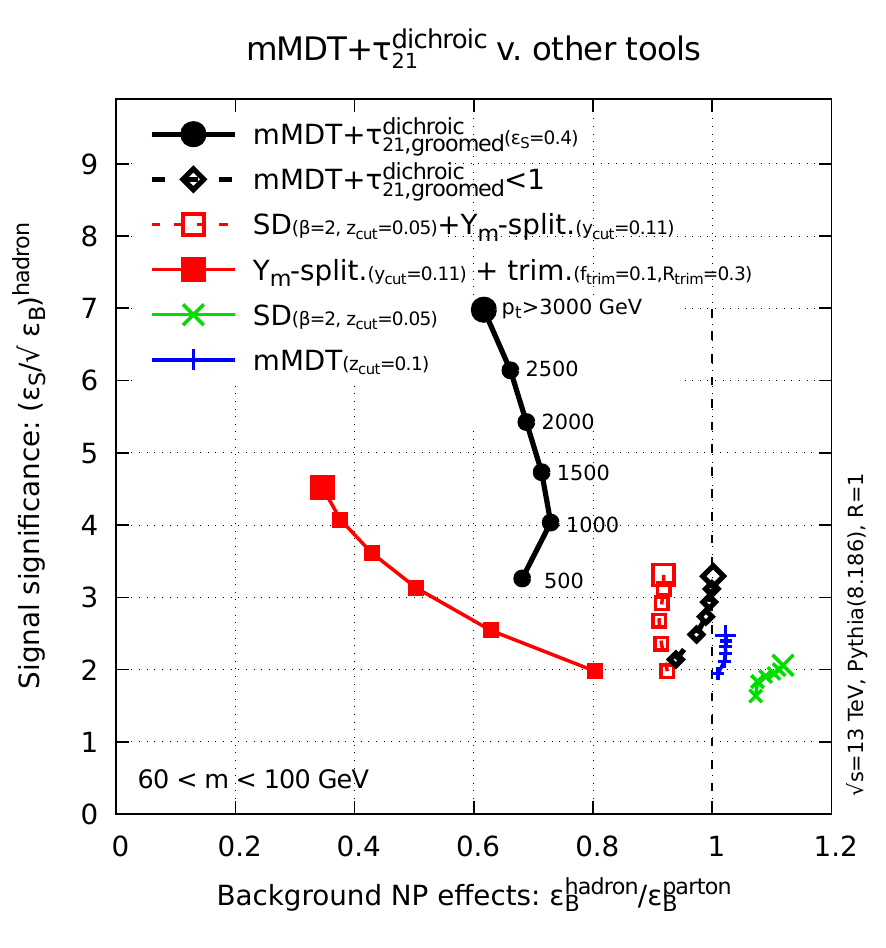}%
\hfill
\includegraphics[width=0.49\textwidth,page=2]{figs/summary.pdf}
}
\caption{
  Signal significance and non-perturbative effects for background, for
  jet $p_t$ cuts ranging from $500\GeV$ to $3\TeV$ in steps of
  $500\GeV$, as in Fig.~\ref{fig:np-effects}(right).
  The $3\TeV$ point is always labelled with a larger symbol.
  The plots compare $\tauDG$ ($\beta_\tau=2$) with a range of other
  tools, including Y$_\text{m}$-splitter (left) and $\beta_\tau=1$
  dichroic subjettiness ratios (right).
  Where the $\beta_\tau$ value is not explicitly labelled, it is equal
  to $2$.
  Note that the default signal-efficiency working point for the
  dichroic subjettiness ratios is $0.4$ here rather than the $0.5$
  chosen in Fig.~\ref{fig:np-effects}. 
  The signal efficiencies for other cases are given in
  Table~\ref{tab:efficiencies}. 
}\label{fig:ccl}
\end{figure}

To complete our Monte Carlo studies, in Fig.~\ref{fig:ccl} we compare
the performance of $\tauDG$ with various other tools:
mMDT tagging alone, SoftDrop grooming alone ($\beta=2$ as above), and also
the Y$_\text{m}$ variant~\cite{ourYsplitter} of
Y-splitter~\cite{Butterworth:2002tt}, combined either with SoftDrop
(pre-)grooming or with trimming, as described in detail in
Ref.~\cite{ourYsplitter} (see also
Ref.~\cite{Dasgupta:2015yua}).
Whereas in the analogous Fig.~\ref{fig:np-effects}, all curves
involved the same signal efficiency, here this is no longer the case.
Accordingly efficiencies are reported versus $p_t$ in
table~\ref{tab:efficiencies}.

\begin{table}[p]
\centering
\begin{tabular}{|l|cccccc|}
\hline
& \multicolumn{6}{c|}{jet $p_t$ cut [GeV]} \\
\cline{2-7}
method &  500 & 1000 & 1500 & 2000 & 2500 & 3000 \\
\hline
mMDT 
 & 0.63 & 0.62 & 0.62 & 0.63 & 0.64 & 0.65 \\
SoftDrop
 & 0.74 & 0.74 & 0.75 & 0.76 & 0.77 & 0.79 \\
Y$_\text{m}$-splitter+trimming
 & 0.49 & 0.41 & 0.36 & 0.33 & 0.31 & 0.30 \\
SoftDrop+Y$_\text{m}$-splitter
 & 0.56 & 0.55 & 0.55 & 0.55 & 0.57 & 0.58 \\
mMDT + $\tauDG<1$
 & 0.60 & 0.57 & 0.58 & 0.58 & 0.59 & 0.61 \\
all other variants
 & 0.4  & 0.4  & 0.4  & 0.4  & 0.4  & 0.4  \\
\hline
\end{tabular}
\caption{Signal efficiencies for the various tools shown in Fig.~\ref{fig:ccl}.}
\label{tab:efficiencies}
\end{table}

Let us start by examining the pure mMDT result: as known already from
\cite{Dasgupta:2013ihk} it provides mild tagging, it has small
non-perturbative corrections and only modest dependence on $p_t$.
SoftDrop ($\beta=2$), when used alone, has slightly lower significance and
larger non-perturbative corrections.\footnote{The performance of SD
  can be somewhat improved for a specific $m/p_t$ value by taking a negative
  value for $\beta$ and adjusting $\zcut$ such that one effectively
  removes branchings with $z<0.1$ at that $m/p_t$ scale (see Section~7
  of Ref.~\cite{Larkoski:2014wba}).}
These two tools have the highest signal efficiencies, of about $63\%$
and $76\%$ respectively at $2\TeV$.

Next we examine combinations that involve Y$_\text{m}$-splitter.
Recall that this tool undoes the last clustering of a
generalised-$k_t$($p=\frac12$) clustering of the jet constituents,
determines $y = \min(p_{t1}^2,p_{t2}^2)\Delta R_{12}^2/m^2$ on the two
resulting prongs, and then imposes a cut $y>y_\text{cut}$.
This cut is similar in its effect to $\zcut$ in mMDT.
When used in conjunction with SD (pre-)grooming, the highest-mass
emission that passes the SD cut is also the one that is unclustered by
Y$_\text{m}$-splitter and so it is required to pass the $y_\text{cut}$
condition.
As a result, the constraint in the Lund plane turns out, at the
leading-log level, to be identical to that obtained with $\tauDG$ and
the condition $\taucut=1$, with a Sudakov suppression vetoing all
emission down to a mass scale $\rho$ in the SD-groomed jet, and a
small prefactor $\sim \alpha_s\ln(1/y_\text{cut})$.
This is reflected in Fig.~\ref{fig:ccl}, where one sees that the
$\tauDG < 1$ curve (black open diamonds) is remarkably similar to the
SD+Y$_\text{m}$-splitter curve (red open squares).
Where the $\tauDG$ variable has an advantage is that one can now
further adjust the choice $\taucut$, whereas with SD+Y$_\text{m}$-splitter
that freedom is not available.

Of the various Y$_\text{m}$ combination considered in
Ref.~\cite{ourYsplitter}, the one that gave the best
signal-to-background discrimination was Y$_\text{m}$ with trimming,
shown as red solid squares in Fig.~\ref{fig:ccl}. 
Overall it performs less well than the mMDT plus $\tauDG$ combination
with $\epsilon_S$ fixed to $0.4$, even though is has a broadly similar
signal efficiency. 

Another point to discuss concerns the choice of
$\beta_\tau$ in the $N$-subjettiness definition, Eq.~(\ref{eq:tauN}).
Many experimental uses of $N$-subjettiness ratios have concentrated on
the choice $\beta_\tau = 1$, while throughout this article we have
used $\beta_\tau = 2$.
A  discussion of the $\beta_\tau=1$ case is given in
Appendix~\ref{app:beta1}, including comparisons of dichroic and normal
variants.
Dichroic always perform best also for $\beta_\tau=1$, and so in the
brief summary that we give here we will only show dichroic results.

An argument often given for the choice of $\beta_\tau = 1$ is that it
is less sensitive to non-perturbative effects.
Fig.~\ref{fig:ccl} (right) shows groomed (filled symbols, solid lines)
and ungroomed (open symbols, dashed lines) results for $\beta_\tau=1$
(squares and triangles) and $\beta_\tau=2$ (circles).
For the $\beta_\tau=1$ case, we have considered either exclusive-$k_t$
axes with the standard $E$-scheme four-vector recombination
(triangles), or the exclusive-$k_t$ axes with the winner-takes-all
(WTA) recombination scheme (squares)~\cite{Bertolini:2013iqa,Larkoski:2014uqa,GPSUnpublished}.
In both the SD-groomed and ungroomed cases, the non-perturbative
corrections are somewhat smaller for $\beta_\tau=1$ (except in the WTA
groomed case).
In the ungroomed case, $\beta_\tau=1$ also leads to better
signal-discrimination.
However once SD-grooming is included the signal discrimination is best
for the $\beta_\tau=2$ case.
If one is concerned about the slightly larger non-perturbative effects
for the SD-groomed $\beta_\tau=2$ case, then one can slightly increase
the $\taucut$ choice: in Fig.~\ref{fig:np-effects}(right) where
$\taucut$ was chosen so as to obtain a higher signal efficiency of
$\epsilon_S=0.5$ the $\tauDG$($\beta_\tau=2$) performance is very
similar to the
$\tauDG$($\beta_\tau=1$,$\epsilon_S=0.4$) performance in
Fig.~\ref{fig:ccl}(right).
Therefore, it is the SD-groomed, $\beta_\tau=2$, dichroic ratio that
appears to give the best overall performance.

There are a number of other variables that one might also consider,
notably energy-correlation functions
(ECFs)~\cite{Larkoski:2013eya,Larkoski:2014gra,Moult:2016cvt}.
In particular we expect that dichroic ratios may be of use also for
the most recent set of ECFs discussed in Ref.~\cite{Moult:2016cvt}, a
number of which are designed to have similarities to $N$-subjettiness.
Their study is, however, beyond the scope of this work.

\section{Brief analytic calculations}\label{sec:analytic}

\begin{figure}
\includegraphics[width=0.3\textwidth]{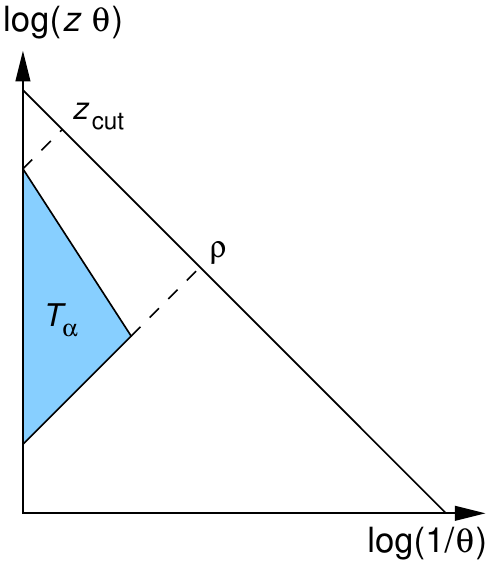}
\hfill
\includegraphics[width=0.3\textwidth]{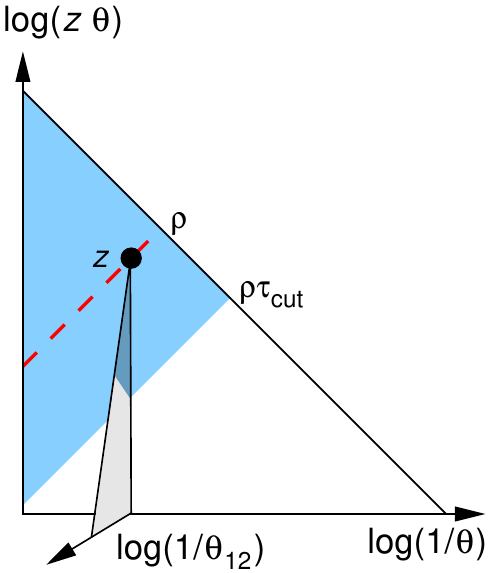}
\hfill
\includegraphics[width=0.3\textwidth]{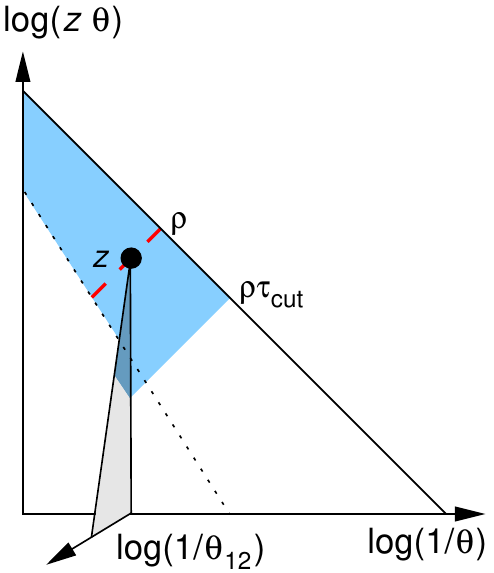}
\caption{Lund diagrams associated with various analytic calculations.
  Left: the basic building block $T_\alpha$,
  Eq.~(\ref{eq:basic-analytic-block}), used to write all Sudakov
  exponents.
  Centre: representation of the \fulljet jet Sudakov
  $R_{\text{\fulljet}}(\rho,\taucut,z)$, Eq.~(\ref{eq:finalRplain}),
  including secondary emissions.
  Right: representation of the \fulljet jet Sudakov
  $R_{\text{SD}}(\rho,\taucut,z)$, Eq.~(\ref{eq:finalRsd}),
  including secondary emissions.
  For both the centre and right plots, the dot indicated by $z$
  corresponds to the emission dominating the jet mass and we will
  integrate over allowed values of its momentum fraction $z$.
}\label{fig:lund-analytic}
\end{figure}

In this Section, we consider brief analytic calculations relating
to the observables we have presented so far.
Our main goal here is to illustrate that the discussion from
Section~\ref{sec:new} --- where we used Lund diagrams to motivate
dichroic subjettiness ratios --- does indeed capture the qualitative
picture observed in Monte-Carlo simulations.
To that aim, it is sufficient to use leading-logarithmic
accuracy, where we control double logarithms, \ie 
$\alpha_s^n \ln^j\! \rho \ln^k\! \rho \ln^\ell\! {\zcut} \ln^m\! {\zetacut}$ 
with $j+k+\ell+m=2n$, assuming
$\rho,\taucut,\zcut,\zetacut \ll 1$. 
Note that, recently, several jet substructure methods have been
understood at higher accuracy, see
\eg \cite{Frye:2016aiz,Larkoski:2015kga}, and we intend to provide a
more precise calculation in future work~\cite{tau21-calculation}.

In practice, we will express everything in terms of the following
fundamental block (cf.\ Fig.~\ref{fig:lund-analytic}(left)):
\begin{equation}\label{eq:basic-analytic-block}
T_\alpha(\rho,\zcut;C_R)
 = \int_0^1\frac{d\theta^2}{\theta^2}\int_0^1\frac{dz}{z}
   \frac{\alpha_s(z\theta p_tR)C_R}{\pi}\Theta(z\theta^2>\rho)
   \Theta(z<\zcut \theta^\alpha)\Theta(\rho<\zcut)\,,
\end{equation}
where angles are normalised to the jet radius $R$ and we use
the 1-loop running-coupling prescription,
$\alpha_s(z\theta p_tR)=\alpha_s/(1+2\alpha_s\beta_0\log z\theta)$
with $\alpha_s\equiv\alpha_s(p_tR)$ and
$\beta_0=(11C_A-4n_fT_R)/(12\pi)$.
Explicit expressions for $T_\alpha$ are given in
Appendix~\ref{app:analytic-blocks} and are mostly taken from
Ref.~\cite{Dasgupta:2015lxh}.
Note that $T_\alpha(\rho,\zcut;C_R)=0$ for $\zcut<\rho$.

For the QCD background, we find, for $\taucut<1$:
\begin{subequations}\label{eq:analytic-all}
\begin{align}
\label{eq:analytic-plain-plain-plain}
\rho_{\text{\fulljet}},\:\tauP:
&& \!\!\left.\frac{\rho}{\sigma}\frac{d\sigma}{d\rho}\right|_{<\taucut}\!
 & \!= \int_\rho^{b_i}\frac{dz}{z} \frac{\alpha_s(\sqrt{z\rho}p_tR)C_R}{\pi}
     \exp\big[\!-\!R_{\text{\fulljet}}(\rho,\taucut,z)\big],\\
\label{eq:analytic-mmdt-mmdt-mmdt}
\rho_{\text{mMDT}},\:\tauT:
&& \!\!\left.\frac{\rho}{\sigma}\frac{d\sigma}{d\rho}\right|_{<\taucut}\!
 & \!= \int_{z_{\text{cut}}}^{b_i}\frac{dz}{z} \frac{\alpha_s(\sqrt{z\rho}p_tR)C_R}{\pi}
     \exp\big[\!-\!R_{\text{mMDT}}(\rho,\taucut,z)\big],\\
\label{eq:analytic-sd-sd-sd}
\rho_{\text{SD}},\:\tauPG:
&& \!\!\left.\frac{\rho}{\sigma}\frac{d\sigma}{d\rho}\right|_{<\taucut}\!
 & \!= \int_{z_{\text{SD}}(\rho)}^{b_i}\frac{dz}{z} \frac{\alpha_s(\sqrt{z\rho}p_tR)C_R}{\pi}
     \exp\big[\!-\!R_{\text{SD}}(\rho,\taucut,z)\big],\\
\label{eq:analytic-mmdt-mmdt-plain}
\rho_{\text{mMDT}},\:\tauD:
&& \!\!\left.\frac{\rho}{\sigma}\frac{d\sigma}{d\rho}\right|_{<\taucut}\!
 & \!= \int_{\zcut}^{b_i}\frac{dz}{z} \frac{\alpha_s(\sqrt{z\rho}p_tR)C_R}{\pi}
     \exp\big[\!-\!R_{\text{\fulljet}}(\rho,\taucut,z)\big],\\
\label{eq:analytic-mmdt-mmdt-sd}
\rho_{\text{mMDT}},\:\tauDG:
&& \!\!\left.\frac{\rho}{\sigma}\frac{d\sigma}{d\rho}\right|_{<\taucut}\!
 & \!= \int_{\zcut}^{b_i}\frac{dz}{z} \frac{\alpha_s(\sqrt{z\rho}p_tR)C_R}{\pi}
     \exp\big[\!-\!R_{\text{SD}}(\rho,\taucut,z)\big],
\end{align}
\end{subequations}
where
$z_{\text{SD}}(\rho)={\text{max}}\big((\rho^\beta\zetacut^2)^{1/(2+\beta)},\rho\big)$
and (cf.\ Fig.~\ref{fig:lund-analytic}(middle,right))
\begin{subequations}\label{eq:finalR}
\begin{align}
R_{\text{\fulljet}}(\rho,\tau,z) 
  & = T_0(\tau\rho,b_i;C_R)
    + T_0(\sqrt{z\rho}\tau,\sqrt{z\rho}b_g;C_A),\label{eq:finalRplain}\\
R_{\text{mMDT}}(\rho,\tau,z) 
  & = R_{\text{\fulljet}}(\rho,\tau,z) 
    - T_0(\tau\rho,\zcut;C_R)
    + T_0(\sqrt{z \rho} \tau, \zcut \sqrt{\rho/z};C_R),\label{eq:finalRmmdt}\\
R_{\text{SD}}(\rho,\tau,z) 
  & = R_{\text{\fulljet}}(\rho,\tau,z) 
    - T_\beta(\tau\rho,\zetacut;C_R)
    + T_\beta(\sqrt{z \rho} \tau, \zetacut (\rho/z)^{(\beta+1)/2};C_R).\label{eq:finalRsd}
\end{align}
\end{subequations}
Note that the \fulljet and mMDT jet mass Sudakov introduced
respectively in Eq.~(\ref{eq:sudakov-rho}) and
Eq.~(\ref{eq:result-mmdtmass}) (and used below) can be written as
\begin{subequations}\label{eq:massR}
\begin{align}
R_{\text{\fulljet}}(\rho) & = R_{\text{\fulljet}}(\rho,1,\text{``any
                            }z\text{''})\,,
\label{eq:Rfull-explicit}
\\
R_{\text{mMDT}}(\rho) & = R_{\text{mMDT}}(\rho,1,\text{``any }z\text{''})\,.
\label{eq:RmMDT-explicit}
\end{align}
\end{subequations}
In the above expressions, $z$ corresponds to the momentum fraction of
the emission dominating the jet mass (emission ``a'' in
Figs.~\ref{fig:combination-3cases} and
\ref{fig:combination-3lund}). Compared to the simple $R'$ factor that
we had in Section~\ref{sec:lund}, we keep the $z$ integration explicit
since the secondary emissions, the $C_A$ terms, depend explicitly on
$z$.
In all cases, the integration over $z$ runs over the region
kinematically allowed by the tagger defining the jet mass.
The Sudakov exponent in these expressions is then essentially given by
the jet on which we compute $\tau_2$.

While we only target leading-logarithmic accuracy, our results also
include the single-logarithmic contributions coming from hard
collinear splittings, which are often phenomenologically important.
They appear as the $b_i$ factors in Eqs.~(\ref{eq:analytic-all}) and
(\ref{eq:finalR}), where we have introduced $b_i=\exp(B_i)$ with
$B_q=-3/4$ and $B_g=-(11C_A-4n_fT_R)/(12C_A)$.
These contributions can effectively be taken into account by limiting
all $z$ integrations to $b_i$ for primary emissions and $b_g$ for
secondary emissions.

Finally, as expected, if one takes the limit $\beta\to \infty$ of the
SD results, one recovers the \fulljet results. Also, the limit
$\beta\to 0$ of (\ref{eq:analytic-sd-sd-sd}), reduces to
(\ref{eq:analytic-mmdt-mmdt-mmdt}).

So far, we have not yet discussed the case where $\rho$ is computed
from the mMDT-tagged jet and $\tau_{21}$ from the \fulljet jet. This is more
involved due to the two separate kinematic configurations involved
(see Fig.~\ref{fig:combination-3cases}(b-c)). In the end, we find
(assuming $\rho<\zcut$)
\begin{multline}
\rho_{\text{mMDT}},\:\tauP:
 \left.\frac{\rho}{\sigma}\frac{d\sigma}{d\rho}\right|_{<\taucut}
 = \int_{\zcut}^{b_i}\frac{dz}{z} \frac{\alpha_s(\sqrt{z\rho}p_tR)C_R}{\pi}
     \exp\big[-R_{\text{\fulljet}}(\rho,\taucut,z)\big]\\
+ \Theta\left(\zcut > \frac{\rho}{\taucut}\right)\int_{\zcut}^{b_i}\frac{dz}{z}\frac{\alpha_s(\sqrt{z\rho}p_tR)C_R}{\pi}
    \exp\big[-R_{\text{mMDT}}(\rho)\big] \times \label{eq:analytic-mmdt-plain-plain}\\
    \times\int_{\rho/\taucut}^{\zcut} \frac{d\rho_c}{\rho_c}
    \int_{\rho_c}^{\zcut}\frac{dz_c}{z_c}\frac{\alpha_s(\sqrt{z_c\rho_c}p_tR)C_R}{\pi}
    \exp\big[-R_{\text{out,\fulljet}}(\rho_c,\taucut,z_c)\big],
\end{multline}
and a similar expression with ``\fulljet'' replaced by ``SD'' for the
case where $\tau_{21}$ is calculated on the SD jet. In the above
expression, we have used $\rho_c=z_c\theta_c^2$ and
\begin{subequations}
\begin{align}
R_{\text{out,\fulljet}}(\rho_c,\tau,z_c)
  & = T_0(\rho_c\tau,\zcut;C_R)
    + T_0(\sqrt{\rho_cz_c}\tau,\sqrt{\rho_cz_c}b_g;C_A)\,,\\
R_{\text{out,SD}}(\rho_c,\tau,z_c)
  & = R_{\text{out,\fulljet}}(\rho_c,\tau,z_c) \nonumber \\
  & - T_\beta(\rho_c\tau,\zetacut;C_R)
    + T_\beta(\sqrt{\rho_cz_c}\tau,\zetacut(\rho_c/z_c)^{(\beta+1)/2};C_R).
\end{align}
\end{subequations}

The configurations contributing to the last two lines of
Eq.~(\ref{eq:analytic-mmdt-plain-plain}) come from jets with at least
one emission in region C (discarded by mMDT) with
$\rho_c\equiv z_c\theta_c^2>\rho/\taucut$.
They result in 
an extra contribution to the mass distribution, which would then be
larger than what we obtain with our dichroic combination
(Eq.~(\ref{eq:analytic-mmdt-mmdt-plain}) or, equivalently, the first
line of Eq.~(\ref{eq:analytic-mmdt-plain-plain})).
When using the dichroic combination, these configurations would all have
$\tau_{21}\ge 1$ (up to $\tau_{21}=\zcut/\rho$). 
In particular, for a cut $\tau_{21}<\taucut$
with $\taucut> 1$, the dichroic combination leads to:
\begin{align}
\label{eq:analytic-mmdt-mmdt-plain-largetau}
&\rho_{\text{mMDT}},\:\tauD:
 \left.\frac{\rho}{\sigma}\frac{d\sigma}{d\rho}\right|_{<\taucut}
 \overset{\taucut> 1}{=}
   \int_{\zcut}^{b_i}\frac{dz}{z} \frac{\alpha_s(\sqrt{z\rho}p_tR)C_R}{\pi}
    \exp\big[-R_{\text{mMDT}}(\rho)\big]\\
& \bigg(\exp\big[-R_{\text{out,\fulljet}}(\rho\taucut)\big]
 + \int_{\rho\taucut}^{\zcut} \frac{d\rho_c}{\rho_c}
    \int_{\rho_c}^{\zcut}\frac{dz_c}{z_c}\frac{\alpha_s(\sqrt{z_c\rho_c}p_tR)C_R}{\pi}
    \exp\big[-R_{\text{out,\fulljet}}(\rho_c,\rho\taucut/\rho_c,z_c)\big]\bigg)\nonumber
\end{align}
with 
\begin{subequations}
\begin{align}
R_{\text{out,\fulljet}}(\rho\tau)  & = T_0(\rho\tau,\zcut;C_R)\\
R_{\text{out,SD}}(\rho\tau) & = R_{\text{out,\fulljet}}(\rho\tau) - T_\beta(\rho\tau,\zetacut;C_R).
\end{align}
\end{subequations}
This result splits into 2 contributions corresponding to the two terms
in the round bracket on the second line of
(\ref{eq:analytic-mmdt-mmdt-plain-largetau}):
the first term comes from configurations where there is no emission in
region C with $z\theta^2>\rho\taucut$, and it corresponds to values of
$\tauD < 1$ (this is manifest, because in
Eq.~(\ref{eq:analytic-mmdt-mmdt-plain-largetau}), given for $\taucut >
1$, it has no dependence on $\taucut$).
For the second contribution, the part corresponding to values of
$\tauD\ge 1$, there is an emission ``c'' with
$z_c\theta_c^2>\rho\taucut$.
To guarantee
$\tau_{21}<\taucut$, we then need to veto emissions (both primary and
secondary) with $z\theta^2>\rho\taucut$.\footnote{Note that the
  difference between the Sudakov suppression in the two contributions
  comes from secondary emissions, i.e. we have
  $R_{\text{out,\fulljet}}(\rho_c,\rho\taucut/\rho_c,z_c) =
  R_{\text{out,\fulljet}}(\rho\taucut) +
  T_0(\sqrt{\rho_cz_c}\taucut,\sqrt{\rho_cz_c}b_g;C_A)$.}
Note that this second contribution itself includes two
sub-contributions: 
the case where emission ``c'' is the only emission
in region C with $z\theta^2>\rho$, yielding a contribution to the
$\tau_{21}$ distribution proportional to $\delta(\tau_{21}-1)$ (recall
that $\tau_2^\text{full}$ is set by the second hardest emission
overall, which makes it equal to $\tau_1^\text{tagged}$); 
and a second sub-contribution where, in addition to emission ``c'',
there is at least one additional 
emission with $\rho\taucut>z\theta^2>\rho$, yielding a continuum with
$\tau_{21}>1$ in the $\tau_{21}$ distribution (see
Fig.~\ref{fig:distribs-tau21} as well as the right plot of
Fig.~\ref{fig:analytic-distribs} below).
One can calculate the $\delta(\tau_{21}-1)$ contribution to the
$\tau_{21}$ distribution by taking the difference between
(\ref{eq:analytic-mmdt-mmdt-plain-largetau}) and
(\ref{eq:analytic-mmdt-mmdt-plain}) for $\taucut\to 1$ which gives
\begin{equation}\label{eq:mixed-delta-tau-1}
\int_{\zcut}^{b_i}\frac{dz}{z} \frac{\alpha_s(\sqrt{z\rho}p_tR)C_R}{\pi}
    \int_{\rho}^{\zcut} \frac{d\rho_c}{\rho_c}
    \int_{\rho_c}^{\zcut}\frac{dz_c}{z_c}\frac{\alpha_s(\sqrt{z_c\rho_c}p_tR)C_R}{\pi}
    e^{-R_{\text{\fulljet}}(\rho)-R_{C_A}(\rho_c,z_c,\rho)},
\end{equation}
with $R_{\text{\fulljet}}(\rho)$ the \fulljet jet mass Sudakov,
Eq.~(\ref{eq:Rfull-explicit}), and
$R_{C_A}(\rho_c,z_c,\rho)=T_0(\sqrt{z_c/\rho_c}\rho,\sqrt{\rho_cz_c}b_g;C_A)$.
Eq.~(\ref{eq:mixed-delta-tau-1}) is equal to the $\taucut \to 1$ limit
of the second term in round brackets in
Eq.~(\ref{eq:analytic-mmdt-mmdt-plain-largetau}). 
In practice the $\delta$-function contribution gets smeared out to
values of $\tau_{12}>1$ through the effect of multiple emissions.

Note that it is relatively straightforward to check that the
limit $\taucut\to 1$ in Eq.~(\ref{eq:analytic-mmdt-plain-plain}), or the
limit $\taucut\to \zcut/\rho$ in
Eq.~(\ref{eq:analytic-mmdt-mmdt-plain-largetau}) both tend to the mMDT jet
mass distribution.

From the equations above, the $\tau_{21}$ distribution, for a given
jet mass, can be obtained by taking the derivative with respect to $\taucut$ and
normalising by the jet mass distribution without any cut on
$\tau_{21}$.
Background efficiencies can also be obtained straightforwardly by
integrating any of the above mass distributions over the allowed mass
window.

For signal jets, we assume that if the jet mass is not within some
reasonable window around the boson mass, then the jet is discarded.
We then find the following signal efficiency
\begin{equation}
\epsilon_S
 = f_{\text{ISR}}\int_{z_{\text{min}}}^{1-z_{\text{min}}}dz\,p_{\text{sig}}(z)
   \exp\big[-R_{\text{sig}}(\rho,\taucut,z)\big],
   \label{eq:epsilon_s}
\end{equation}
with $z_{\text{min}}=\rho$, $z_{\text{SD}}(\rho)$ or $\zcut$ depending
on whether the mass is computed on the \fulljet jet, the SD-groomed
jet or the mMDT-tagged jet, respectively.
The $\tau_{21}$ distribution for a given jet mass can be obtained by
taking the derivative of $\epsilon_S$ with respect to $\taucut$ (and
normalising appropriately).

In Eq.~(\ref{eq:epsilon_s}) the Sudakov exponent is given by
\begin{align}
R_{\text{sig}}(\rho,\tau,z)
& = \Big[T_0(\sqrt{z(1-z)\rho}\tau;\sqrt{(1-z)\rho/z}b_i;C_R)\nonumber\\
&{\phantom{=}}-T_0(\sqrt{z\rho/(1-z)};\sqrt{(1-z)\rho/z}b_i;C_R)\Big]
  \nonumber\\
& + \Big[T_0(\sqrt{z(1-z)\rho}\tau;\sqrt{z\rho/(1-z)}b_i;C_R)\nonumber\\
&{\phantom{=}}-T_0(\sqrt{(1-z)\rho/z};\sqrt{z\rho/(1-z)}b_i;C_R)\Big] ,
\end{align}
valid for small $\tau$.
Here we target double-logarithmic accuracy, $\alpha_s^n\ln^{2n} \tau$,
though we also include a set of finite-$z$ and hard-splitting
corrections that were found to be numerically important in
Ref.~\cite{Dasgupta:2015lxh} (cf.\ Eq.~(A.24)). These represent
only a subset of next-to-leading logarithmic terms.
Note that for $z\ll 1$ ($1-z\ll 1$) the term on the fourth (second)
line is zero because of the last of the $\Theta$-functions in
Eq.~(\ref{eq:basic-analytic-block}), while the term on the third
(first) line corresponds to the leaf in
Fig.~\ref{fig:lund-nsubjettiness}(right). 
For simplicity, in our numerical results we will use $p_{\text{sig}}(z)=1$ in
Eq.~(\ref{eq:epsilon_s}).\footnote{For the $WW$ 
  process under consideration, correlations between the incoming
  quarks and the final quarks after the decay of the two $W$ bosons
  have been calculated in \cite{Gunion:1985mc} and could in principle
  be used to compute $p_{\text{sig}}(z)$. This would however be
  specific to the $WW$ process considered here just as an example. We
  therefore use the ``splitting function'' of an unpolarised $W$
  boson. This simplification does not affect significantly any of the
  results presented here.}

Eq.~(\ref{eq:epsilon_s}) also includes a factor $f_\text{ISR}$ that
accounts for the effect of initial-state radiation (ISR).
Such effects are present both for signal and background jets and are
generically single-logarithmic.
As such they are subleading compared to the double-logarithms that we
resum. 

Nevertheless, if we consider signal jets and examine the limit
of large $p_t$ with $M$, $\taucut$, etc.\ all fixed, then because of
the absence of double logarithms of $\rho$, single-logarithmic ISR
effects $(\as\ln\rho)^n$ can be numerically
dominant~\cite{Dasgupta:2015yua}.
Physically, they are associated with the requirement that ISR should
not substantially modify the mass of the signal jet.
The correction involves $(\as\ln\rho)^n$ terms, only when the mass is
determined on the \fulljet jet and the factor $f_\text{ISR}$ then takes the
form
\begin{align}
f_{\text{ISR}} & =
\exp\Big[-\frac{C_R}{2\pi\beta_0}R^2\,\log\frac{1}{1-2\lambda}\Big],\\
\lambda & = \beta_0\alpha_s(p_t)\left[
          \log \frac{1}{\rho} 
          + {\cal O}\left(\log\frac{M}{\delta M},\log\frac{1}{\taucut},\cdots\right)
          \right],
\end{align}
where a non-global contribution (formally of the same logarithmic order)
is ignored for simplicity.
In the above formula, $\delta M$ is size of the mass window in
which signal jets are accepted, and a full treatment of all
single-logarithmic corrections would need to account also for
logarithms of $\delta M/M$.
A more complete treatment of $f_\text{ISR}$  would be relevant for precise
phenomenological applications.
The finite $\order{\as}$ component associated with high-$p_t$
emissions could be obtained \eg using POWHEG~\cite{powheg},
aMC@NLO~\cite{Alwall:2014hca} or at NNLO using
MATRIX~\cite{deFlorian:2016uhr} or MCFM~\cite{mcfm}.

\begin{figure}[t]
\centering{%
  \includegraphics[width=0.48\textwidth,page=2]{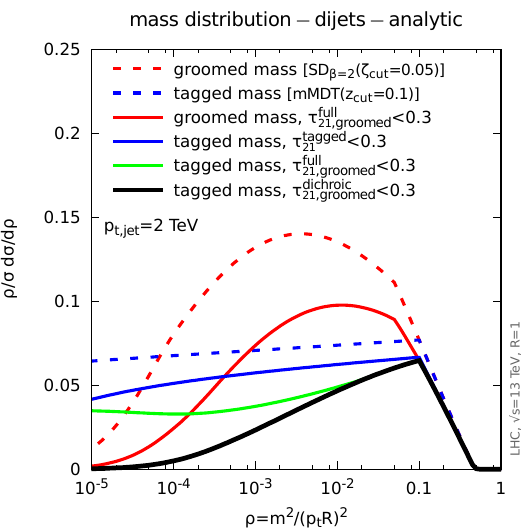}%
  \hfill%
  \includegraphics[width=0.48\textwidth,page=2]{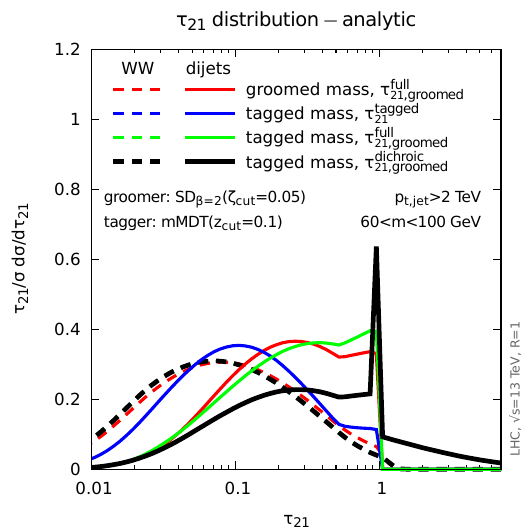}
}
\caption{Same as figure as \ref{fig:distribs-mass} and
  \ref{fig:distribs-tau21} now obtained from our analytic calculation
  instead of Monte-Carlo simulations.
  In the right-hand plot, for clarity, the $\delta$-function that
  appears at $\tauDG=1$ (dijets) has been represented with finite
  width and scaled down by a factor of $5$.
}\label{fig:analytic-distribs}
\end{figure}
\begin{figure}[t]
\centering{%
\includegraphics[width=0.48\textwidth,page=3]{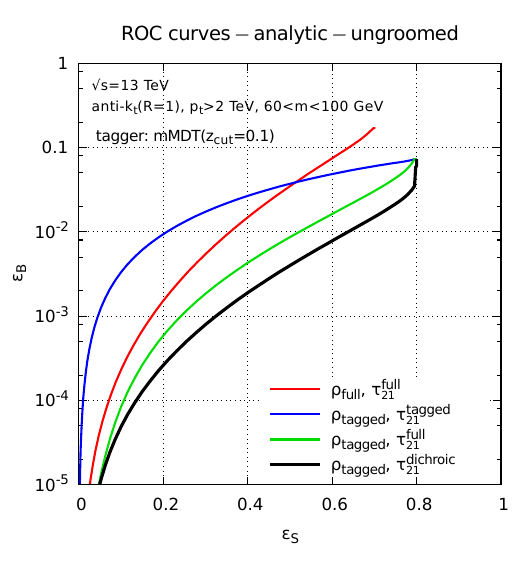}%
\hfill
\includegraphics[width=0.48\textwidth,page=4]{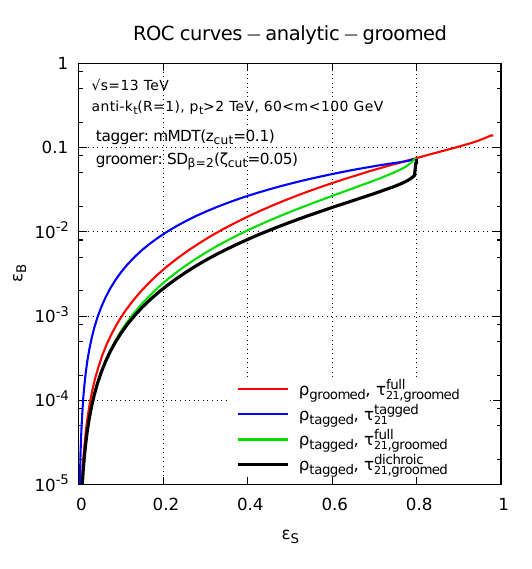}%
}
\caption{Same as figure as \ref{fig:roc-parton} now obtained from our
  analytic calculation instead of Monte-Carlo
  simulations.}\label{fig:analytic-roc}
\end{figure}

We can now compare our analytic predictions with the Monte-Carlo results
from the previous Section.
We  use $\alpha_s(M_Z)=0.1383$, as in the
Pythia~8 simulations presented in the previous Section, and freeze the
coupling for scales below $\mu_{\text{fr}}=\tilde\mu_{\text{fr}}p_tR$,
which we set to 1~GeV.
We start with the QCD mass distributions, shown on the left plot of
Fig.~\ref{fig:analytic-distribs}, to be compared to the Monte-Carlo
results presented in Fig.~\ref{fig:distribs-mass}.
Globally, we see that our analytic calculation captures correctly the
main patterns discussed earlier.
We note however that the analytic distributions, especially those
involving the \fulljet jet mass, are less peaked than the Monte-Carlo
ones. This is likely due to subleading logarithmic corrections, like
multiple-emission corrections which would effectively increase the
Sudakov exponent.

The $\tau_{21}$ distributions for both QCD jets and signal ($W$) jets
are shown in the right plot of Fig.~\ref{fig:analytic-distribs}, to be
compared with Fig.~\ref{fig:distribs-tau21}.
The ordering between the different curves is well captured by our
analytic expressions.
Differences related to the over-simplicity of our leading-logarithmic
approximation are larger than what was seen for the mass
distribution. 
First, our analytic calculations are non-zero when $\tau_{21}\to 1$.
This region is however not under control within our strongly-ordered
approximation.
Similarly, the kink observed for $\tau_{21}\sim 0.5$ is not
physical. It comes from the onset of the secondary-emission
contribution which starts, in our formulas, at $\tau_{21}=b_g$. 
The analytic calculation for our dichroic
combination is given by the black curves in the right plot of
Fig.~\ref{fig:analytic-distribs}. 
The dijet case clearly has a contribution proportional to
$\delta(\tau_{21}-1)$ (cf. Eq.~(\ref{eq:mixed-delta-tau-1})) (scaled
down by a factor of $5$ for clarity), which is not observed in the
Monte-Carlo results.
In practice, additional emissions at smaller $z\theta^2$ would also
contribute to $\tau_{21}$, and they would transform the
$\delta(\tau_{21}-1)$ contribution into a Sudakov peak at
$\tau_{21}\gtrsim 1$, which is visible on the Monte-Carlo simulations.
We are currently working on a better analytic calculation, lifting the
assumption that emissions are strongly ordered in
$z\theta^2$~\cite{tau21-calculation}.

Finally, let us turn to the ROC curves, plotted in
Fig.~\ref{fig:analytic-roc}. We again see that they reproduce
the main qualitative features observed in Section~\ref{sec:mc-study}.
There are however quantitative differences between our analytic
results and the Monte-Carlo simulations. For example, our calculation
over-estimates the signal efficiencies. 
A more quantitative description would require a more precise analytic
treatment including subleading corrections, beyond the strong-ordering
approximation, and fixed-order corrections for signal efficiencies.

\section{Conclusion}\label{sec:ccl}

In this paper we have examined the interplay between boosted-object
tagging algorithms, mMDT or SoftDrop, and radiation constraints,
notably as imposed through $N$-subjettiness cuts.
The analysis points to a new $N$-subjettiness ratio, $\tauD =
\tau_{2}^\text{\fulljet}/\tau_{1}^\text{tagged}$, where the numerator
is evaluated on the \fulljet jet, while the denominator is evaluated
on the set of constituents left after the tagging stage.
The name ``dichroic'' comes from the fact that the large-angle colour
flow, present in backgrounds but not signals, gets directed
exclusively to the numerator and not the denominator.
It is this feature that leads to an enhanced significance in
distinguishing (colour-singlet) signals from (colour-triplet or octet)
backgrounds, notably compared to current widely used $N$-subjettiness
ratios.

As well as considering signal-significance, it is important to keep
non-perturbative effects under control: a method that is overly
reliant on non-perturbative physics for its discrimination power is
one for which signal-efficiency and background-rejection estimates may
be highly model-dependent, and correspondingly uncertain.
It is also likely to be subject to large detector effects.
We have found that the combination of $\tauD$ with a light grooming step
based on SoftDrop ($\beta=2$),
$\tauDG = \tau_{2}^\text{SD}/\tau_{1}^\text{tagged}$ is effective in
maintaining good signal-to-background significance while substantially
limiting non-perturbative effects.

The overall behaviour of our dichroic $\tau_{21}$ variable, with
grooming, was illustrated in Fig.~\ref{fig:distribs-tau21}: the
$\tau_{21}$ distribution for signal jets is left largely unmodified by
the change to a dichroic variant (black dashed curve versus any of the
other dashed curves), whereas the distribution for background jets is
shifted to substantially higher values of $\tau_{21}$ (black solid
curve versus any of the other solid curves), increasing the ability to
distinguish signal and background.

Figures \ref{fig:np-effects} and \ref{fig:ccl} provide a summary of
the signal-significance (vertical axis) and non-perturbative
corrections (horizontal axis) for a range of boosted-object
identification methods.
The points along the lines correspond to different signal-efficiency
working points (Fig.~\ref{fig:np-effects}(left)) or  $p_t$ cuts (the
other plots).
One sees that $\tauDG$ with $\beta_\tau=2$, in black, provides the
best signal significance of any of the methods and that, for a given
signal significance, it tends to limit the size of non-perturbative
effects relative to other methods.

In addition to the Lund-plane based arguments given in
section~\ref{sec:new} and the Monte Carlo studies of
section~\ref{sec:mc-study}, we have also outlined the
analytic leading-logarithmic structure of different combinations of
taggers and $\tau_{21}$ ratios.
As well as bringing insight into the behaviour of different taggers,
such calculations provide a basis for the future design of
``decorrelated'' \cite{Dolen:2016kst} combinations of taggers and
dichroic radiation constraints, providing background rejection that is
independent of the tagged jet mass and thus straightforward to use in the
context of data-driven background estimates.

\section*{Acknowledgements}
GS thanks CERN for hospitality while part of this work was being
finalised.
GS's work is supported in part by the French Agence Nationale de la
Recherche, under grant ANR-15-CE31-0016.
GPS and GS are both supported in part by ERC Advanced Grant
Higgs@LHC (No.\ 321133). 

\appendix

\section{Dichroic subjettiness ratios for $\beta_\tau=1$}\label{app:beta1}

In Section~\ref{sec:new}, we have argued in favour of the dichroic
subjettiness ratios using $N$-subjettiness with $\beta_\tau=2$. 
In this appendix, we briefly discuss the case $\beta_\tau=1$, for
which the dichroic variant can also be considered.
Note that for $\beta_\tau=1$, we have defined the $N$-subjettiness
axes through an exclusive-$k_t$ declustering. This can be done either
using the standard $E$-scheme four-vector recombination or the
winner-takes-all (WTA) recombination scheme.
For simplicity, we will focus on $E$-scheme results here.
A brief comparison between the two axis choices is shown in
Fig.~\ref{fig:ccl}(right).

\begin{figure}
\centerline{
\includegraphics[width=0.48\textwidth,page=1]{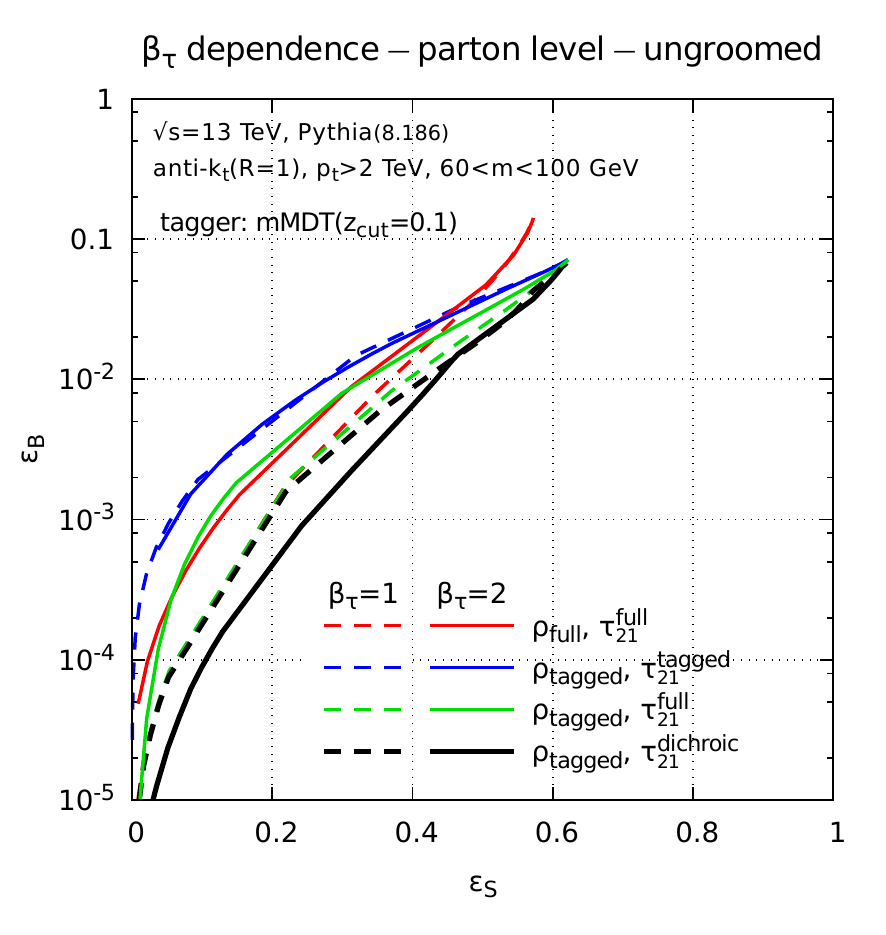}%
\hfill
\includegraphics[width=0.48\textwidth,page=3]{figs/rocs-beta.pdf}%
}
\centerline{
\includegraphics[width=0.48\textwidth,page=2]{figs/rocs-beta.pdf}%
\hfill
\includegraphics[width=0.48\textwidth,page=4]{figs/rocs-beta.pdf}%
}
\caption{ROC curves providing a comparison between different
  $N$-subjettiness ratios for $\beta_\tau=1$ (dashed lines) and
  $\beta_\tau=2$ (solid lines). The same 4 variants as in
  Figs.~\ref{fig:roc-parton} and~\ref{fig:roc-full} are included. The
  left (right) column corresponds to \fulljet (SD-groomed) jets. The
  top (bottom) row corresponds to parton-level (hadron-level)
  events.
}\label{fig:beta1}
\end{figure}

Fig.~\ref{fig:beta1} shows ROC curves similar to those presented in
Figs.~\ref{fig:roc-parton} and~\ref{fig:roc-full}, this time including
results for $\beta_\tau=1$ as dashed lines.

We can make several observations based on these plots.
First, as for $\beta_\tau=2$, we see that the dichroic ratio also
outperforms the other combination for $\beta_\tau=1$. The performance
gain is however smaller, especially with SD grooming.

In terms of the sensitivity to non-perturbative effects, we see that
$N$-subjettiness ratios with $\beta_\tau=1$ are rather stable even
without any SD grooming step. 
This small sensitivity to non-perturbative effects might have been
anticipated since the corresponding $k_t$ cut is less affected by
soft-and-large-angle emissions than for $\beta_\tau=2$.
A consequence of this observation is that grooming is less
critical when using a cut on $N$-subjettiness ratios with
$\beta_\tau=1$, and without SD grooming the dichroic combination
shows a more sizeable performance gain compared to the other
approaches, cf.\ the bottom-left plot of Fig.~\ref{fig:beta1}.

Finally, we can argue that $\beta_\tau=2$ gives somewhat better
performance than $\beta_\tau=1$.
To be fair, the comparison should be made between $\tauDG$ for
$\beta_\tau=2$ (the solid black line on the bottom-right plot of
Fig.~\ref{fig:beta1}) and $\tauD$ for $\beta_\tau=1$ (the dashed black
line on the bottom-left plot) which both show good signal significance
and limited non-perturbative corrections. 
This comparison shows a somewhat larger background rejection in the
$\beta_\tau=2$ case for typical signal efficiencies in the $0.2{-}0.6$
range, as also seen in Fig.~\ref{fig:ccl}.

\section{Explicit expressions for the analytic results}\label{app:analytic-blocks}

For completeness, we give the result of the building block used for
all the analytic calculations in Section~\ref{sec:analytic}, see
Eq.~\eqref{eq:basic-analytic-block}.

We work with a one-loop running coupling (with 5
active flavours), appropriate at our accuracy. 
We take $\alpha_s(M_Z)=0.1383$ to match with our Pythia simulations
and freeze the coupling below a scale
$\mu_{\text{fr}}=\tilde\mu_{\text{fr}}p_tR$ which we set to 1~GeV in
practice.
We then find
\newcommand{\wideeq}{\makebox[{1.5cm}]{$=$}}
\newcommand{\widespace}{\makebox[{1.5cm}]{ }}
\begin{align}
T_\alpha(\rho,\rho_0;C_R)
 & \overset{L<L_{\text{fr}}}{\wideeq}
     \frac{C_R}{2\pi\alpha_s\beta_0^2}\bigg[
     \frac{W(1-\lambda_0)}{1+\alpha}+W(1-\lambda)-\frac{2+\alpha}{1+\alpha}W(1-\bar\lambda)
     \bigg]\\
 & \overset{\bar L<L_{\text{fr}}<L}{\wideeq}
     \frac{C_R}{2\pi\alpha_s\beta_0^2}\bigg[
     \frac{W(1-\lambda_0)}{1+\alpha}+(1-\lambda)\log(1-\lambda_{\text{fr}})-\frac{2+\alpha}{1+\alpha}W(1-\bar\lambda)+\lambda_{\text{fr}}-\lambda
     \bigg]\nonumber\\
 & \widespace +\frac{\alpha_s(\mu_{\text{fr}})C_R}{\pi}(L-L_{\text{fr}})^2\\
 & \overset{L_0<L_{\text{fr}}<\bar L}{\wideeq}
     \frac{C_R}{2\pi\alpha_s\beta_0^2}\frac{1}{1+\alpha}\bigg[
     W(1-\lambda_0)-(1-\lambda_0)\log(1-\lambda_{\text{fr}})+\lambda_0-\lambda_{\text{fr}}
     \bigg]\nonumber\\
 & \widespace  +\frac{\alpha_s(\mu_{\text{fr}})C_R}{\pi}\bigg[
   (L-\bar L)^2 + \frac{1}{1+\alpha}(\bar L-L_{\text{fr}})(\bar L+L_{\text{fr}}-2L_0)\bigg]\\
 & \overset{L_0>L_{\text{fr}}}{\wideeq}
    \frac{\alpha_s(\mu_{\text{fr}})C_R}{\pi}\frac{1}{2+\alpha}(L-L_0)^2,
\end{align}
with $W(x)=x\log(x)$ and 
\begin{align}
L      & = \log(1/\rho),
 &  \lambda    & =2\alpha_s\beta_0L,\\
L_0    & = \log(1/\rho_0),
 &  \lambda_0  & =2\alpha_s\beta_0L_0,\\
L_{\text{fr}}    & = \log(1/\tilde\mu_{\text{fr}}),
 &  \lambda_{\text{fr}} & =2\alpha_s\beta_0L_{\text{fr}},\\
\bar L & = \frac{L_0+(1+\alpha)L}{2+\alpha},
 & \bar\lambda & =2\alpha_s\beta_0 \bar L.
\end{align}

\section{Example code for dichroic subjettiness
  ratios}\label{app:code}

In this last Appendix, we briefly indicate how dichroic subjettiness
ratios can be implemented using tools available in \ttt{FastJet} and
\ttt{fjcontrib}.

First, besides standard \ttt{FastJet} headers needed for jet
clustering, one needs to include the following headers:
\begin{lstlisting}
  #include <fastjet/contrib/ModifiedMassDropTagger.hh> // mMDT tagger
  #include <fastjet/contrib/SoftDrop.hh>               // optional SD grooming
  #include <fastjet/contrib/Nsubjettiness.hh>          // tau1 and tau2
\end{lstlisting}
Then, one should declare the basic objects needed for tagging,
computing $\tau_1$ and $\tau_2$, and, optionally, grooming:
\begin{lstlisting} 
  // the tagger [here mMDT with a z cut]
  // Note: by default, this automatically reclusters with Cambridge/Aachen
  double zcut = 0.1;
  fastjet::contrib::ModifiedMassDropTagger mmdt_tagger(zcut);

  // (optional) groomer [here SoftDrop]
  // Note: by default, this automatically reclusters with Cambridge/Aachen
  double beta    = 2.0;
  double zetacut = 0.05;
  fastjet::contrib::SoftDrop sd_pre_groomer(beta, zetacut);

  // N-subjettiness with beta_tau=2 and gen-kt axes
  double beta_tau = 2.0;
  fastjet::contrib::UnnormalizedMeasure measure(beta_tau);
  fastjet::contrib::GenKT_Axes axes_gkt(1.0/beta_tau);
  fastjet::contrib::Nsubjettiness tau1(1, axes_gkt, measure);
  fastjet::contrib::Nsubjettiness tau2(2, axes_gkt, measure);
\end{lstlisting}
Note that all parameters here are given as examples and have not been
optimised. Also, when used with events contaminated by pileup, a
proper pileup mitigation technique should be implemented. This can for
example be done by passing a \ttt{fastjet::Subtractor} to the mMDT and
SD via the \ttt{set\_subtractor} method, and using a
\ttt{GenericSubtractor}~\cite{Soyez:2012hv} or a
\ttt{ConstituentSubtractor}~\cite{Berta:2014eza} for the
$N$-subjettiness variables.
Alternatively one can use methods that carry out event-wide
pileup-suppression such as PUPPI~\cite{Bertolini:2014bba} or
SoftKiller~\cite{Cacciari:2014gra}.

Finally, for a given jet (\ttt{jet} in the example below), one can
compute the dichroic subjettiness ratio using
\begin{lstlisting} 
  fastjet::PseudoJet jet; // given jet

  fastjet::PseudoJet pre_groomed_jet = sd_pre_groomer(jet);     // grooming
  fastjet::PseudoJet tagged_jet = mmdt_tagger(pre_groomed_jet); // tagging

  double tau1_tagged    = tau1(tagged_jet);          // $\tau_1^{\text{tagged}}$
  double tau2_groomed   = tau2(pre_groomed_jet);     // $\tau_2^{\text{groomed}}$
  
  double tagged_mass = tagged_jet.m();               // tagged mass  
  double tau21_dichroic = tau2_groomed/tau1_tagged;  // $\tauDG$
\end{lstlisting}

\end{document}